\documentclass[aps,prc,showpacs,twocolumn,floatfix,superscriptaddress,nofootinbib]{revtex4-1}

\usepackage{graphicx}
\usepackage{epsfig}
\usepackage{bm}
\usepackage{amsmath}
\usepackage{amssymb}
\usepackage{verbatim}

\begin{document}
\title{Ab initio Bogoliubov coupled cluster theory for open-shell nuclei}

\author{A. Signoracci}
\email{asignora@utk.edu}
\affiliation{Department of Physics and Astronomy, University of Tennessee, Knoxville, TN 37996, USA}
\affiliation{Physics Division, Oak Ridge National Laboratory, Oak Ridge, TN 37831, USA}
\affiliation{CEA-Saclay, IRFU/Service de Physique Nucl\'eaire, F-91191 Gif-sur-Yvette, France}

\author{T. Duguet}
\email{thomas.duguet@cea.fr}
\affiliation{CEA-Saclay, IRFU/Service de Physique Nucl\'eaire, F-91191 Gif-sur-Yvette, France}
\affiliation{National Superconducting Cylcotron Laboratory and Department of Physics and Astronomy, Michigan State University, East Lansing, MI 48824, USA}

\author{G. Hagen}
\email{hageng@ornl.gov}
\affiliation{Physics Division, Oak Ridge National Laboratory, Oak Ridge, TN 37831, USA}
\affiliation{Department of Physics and Astronomy, University of Tennessee, Knoxville, TN 37996, USA}

\author{G.R. Jansen}
\email{gustav.jansen@utk.edu}
\affiliation{Physics Division, Oak Ridge National Laboratory, Oak Ridge, TN 37831, USA}
\affiliation{Department of Physics and Astronomy, University of Tennessee, Knoxville, TN 37996, USA}

%
%
\begin{abstract}
\begin{description}

\item[Background] Ab initio many-body methods have been developed over the past ten years to address closed-shell nuclei up to mass $\text{A}\sim 130$ on the basis of realistic two- and three-nucleon interactions. A current frontier relates to the extension of those many-body methods to the description of open-shell nuclei. 
\item[Purpose] Several routes to address open-shell nuclei are currently under investigation, including ideas which exploit spontaneous symmetry breaking. Singly open-shell nuclei can be efficiently described via the sole breaking of $U(1)$ gauge symmetry associated with particle-number conservation, as a way to account for their superfluid character. While this route was recently followed within the framework of self-consistent Green's function theory, the goal of the present work is to formulate a similar extension within the framework of coupled cluster theory.
\item[Methods] We formulate and apply Bogoliubov coupled cluster (BCC) theory, which consists of representing the exact ground-state wavefunction of the system as the exponential of a quasiparticle excitation cluster operator acting on a Bogoliubov reference state. Equations for the ground-state energy and the cluster amplitudes are derived at the singles and doubles level (BCCSD) both algebraically and diagrammatically. The formalism includes three-nucleon forces at the normal-ordered two-body level. The first BCCSD code is implemented in $m$-scheme, which will permit the treatment of doubly open-shell nuclei via the further breaking of $SU(2)$ symmetry associated with angular momentum conservation.
\item[Results] Proof-of-principle calculations in an $N_{\text{max}}=6$ spherical harmonic oscillator basis are performed for $^{16,18,20}$O, $^{18}$Ne, and $^{20}$Mg in the BCCD approximation with a chiral two-nucleon interaction, comparing to results obtained in standard coupled cluster theory when applicable. The breaking of $U(1)$ symmetry is monitored by computing the variance associated with the particle-number operator.
\item[Conclusions] The newly developed many-body formalism increases the potential span of ab initio calculations based on single-reference coupled cluster techniques tremendously, i.e. potentially to reach several hundred additional mid-mass nuclei. The new formalism offers a wealth of potential applications and further extensions dedicated to the description of ground- and excited-states of open-shell nuclei. Short-term goals include the implementation of three-nucleon forces at the normal-ordered two-body level. Mid-term extensions include the approximate treatment of triple corrections and the development of the equation-of-motion methodology to treat both excited states and odd nuclei. Long-term extensions include exact restoration of $U(1)$ and $SU(2)$ symmetries.
\end{description}

\end{abstract}

\maketitle

%

\section{Introduction}
\label{introduction}

Ab initio many-body methods based on coupled cluster (CC)~\cite{Dean:2003vc,Kowalski:2003hp,Wloch:2005za,Wloch:2005qq,Gour:2005dm,Hagen:2007ew,Hagen:2010gd,Jansen:2011gb,Binder:2012mk,Binder:2013oea}, self-consistent Dyson-Green's function (SC{\it Dy}GF)~\cite{Barbieri:2000pg,Barbieri:2001gt,Dickhoff:2004xx,Waldecker:2011by,Cipollone:2013zma} and in-medium similarity renormalization group (IMSRG)~\cite{Tsukiyama:2010rj,Hergert:2012nb} techniques have been intensively developed in the last ten years to address nuclei up to mass $\text{A}\sim 130$~\cite{Binder:2013xaa}. However, these important developments have been limited until recently to doubly closed-(sub)shell nuclei plus those accessible via the addition and removal of one or two nucleons. 

Extending many-body methods to genuinely open-shell nuclei necessarily complicates the formalism and increases the computational cost. One possible way to overcome the near degeneracy of the reference state relies on the development of multi-reference (MR) methods. Recently, a multi-reference IMSRG technique has been formulated and implemented to address (singly) open-shell nuclei~\cite{Hergert:2013uja} whereas CC-based~\cite{Jansen:2014qxa} and IMSRG-based~\cite{Bogner:2014baa} configuration interaction methods have been proposed as well. 

An alternative route exploits the concept of spontaneous symmetry breaking, where $U(1)$ gauge symmetry associated with particle-number conservation can be broken to capture the superfluid character of singly open-shell nuclei in a controlled manner. Addressing doubly open-shell systems relies on the breaking of another symmetry, i.e. $SU(2)$ rotational symmetry associated with angular momentum conservation, to grasp quadrupole correlations. The breaking of $U(1)$ symmetry has been recently exploited within the framework of Green's function techniques via the first ab initio application of self-consistent Gorkov-Green's function (SC{\it Go}GF) theory to finite nuclei~\cite{soma11a,Soma:2012zd,Barbieri:2012rd,Soma:2013vca}. First results in the calcium region based on realistic two- and three-nucleon chiral forces show great promise~\cite{Soma:2013xha}.

In this context, the goal of the present work is to extend single-reference CC theory in a way that allows for the breaking of $U(1)$ symmetry. We formulate a workable Bogoliubov coupled cluster (BCC) theory for nuclei by representing the exact ground-state wavefunction of even-even open-shell nuclei as the exponential of a quasiparticle excitation cluster operator acting on a Bogoliubov reference state in order to extend the reach of single-reference coupled cluster calculations~\cite{stolarczyk2010}. A reduced form of this theory based on a Bardeen-Cooper-Schrieffer (BCS) reference state was already formulated and applied to simplified, e.g. translationally invariant, geometries~\cite{emrich84a,lahoz88a}. Very recently, the BCS-based version of the BCC formalism developed in the present paper was applied, at the doubles level, to the attractive pairing Hamiltonian problem~\cite{Henderson:2014vka}. Near the transition point where particle-number symmetry is spontaneously broken, a high-quality reproduction of exact Richardson solutions~\cite{richardson63a,richardson64a} was obtained. The present work derives BCC theory and, encouraged by the results of \textcite{Henderson:2014vka}, applies it for the first time to ab initio calculations of open-shell nuclei. 

The paper is organized as follows. Sections~\ref{setup} and~\ref{s:bcc} formulate the general BCC theory before providing fully expanded expressions of the equations at the singles and doubles (BCCSD) level in Sec.~\ref{BCCSDsection}. The diagrammatic method for the BCC formalism, as well as the full set of diagrams at play at the BCCSD level, is treated in Sec.~\ref{diagrammatic}. Results of the first proof-of-principle calculations are discussed in Sec.~\ref{applications}. Conclusions are given in Sec.~\ref{conclusions}, while two appendices provide additional technical details.

\section{Bogoliubov setting}
\label{setup}

\subsection{Hamiltonian}

The nuclear Hamiltonian $H=T_{\text{kin}}+V+W$ is the sum of the kinetic energy operator and of internucleon interactions truncated at the three-body level. The Hamiltonian can be expressed in an arbitrary single-particle basis under the second-quantized form
\label{e:ham}
\begin{eqnarray}
\nonumber H &\equiv& \,\,\,\,\, \frac{1}{(1!)^2} \sum _{pq} t_{pq} c^{\dagger}_{p} c_{q}+\frac{1}{(2!)^2} \sum _{pqrs} \bar{v}_{pqrs}  c^{\dagger}_{p} c^{\dagger}_{q} c_{s} c_{r} \\
  &&       + \frac{1}{(3!)^2} \sum_{pqrstu} \bar{w}_{pqrstu} c^{\dagger}_{p}c^{\dagger}_{q}c^{\dagger}_{r}c_{u}c_{t}c_{s} \, ,
\end{eqnarray} 
employing antisymmetric matrix elements of two- and three-body interactions.

As self-bound systems, the center-of-mass motion of nuclei can be separated from the motion of the nucleons relative to it.\footnote{This separation was demonstrated in practical CC applications~\cite{Hagen:2010gd,Jansen:2013gb}, while its verification in the BCC framework will be a subject of future investigation.}  Being interested in the intrinsic energy of the system, we subtract the center-of-mass contribution to the Hamiltonian
\begin{equation}
\label{eq:H_in}
H_{\text{rel}} = H - H_{\text{cm}} = T^{\text{1B}}_{\text{rel}} + [V+T^{\text{2B}}_{\text{rel}}] + W \, ,
\end{equation}
where the relative kinetic energy was decomposed into one- and two-body contributions defined respectively as 
\begin{subequations}
\label{relativeT}
\begin{eqnarray}
T^{\text{1B}}_{\text{rel}} &\equiv& \left( 1- \frac{1}{\text{A}} \right ) \sum_i \frac{\mathbf{p}_i^2}{2M} \, , \label{relativeT1B} \\
T^{\text{2B}}_{\text{rel}} &\equiv& - \frac{1}{\text{A}} \sum_{i<j} \frac{\mathbf{p}_i \cdot \mathbf{p}_j}{M} \, , \label{relativeT2B}
\end{eqnarray}
\end{subequations}
with $\mathbf{p}_i $ the momentum of the $i$-th nucleon, $M$ the nucleon mass and $\text{A}$ the number of nucleons. In Eq. \eqref{relativeT}, the term $1/\text{A}$ should really be seen as the inverse of the particle-number {\it operator} $A$. While it can be straightforwardly replaced by the {\it number} $\text{A}$ in particle-number-conserving theories, it is not the case for the BCC scheme developed here once the many-body expansion is truncated, as good particle number is then only conserved on average. It could however be shown~\cite{Hergert:2009na} that the form given in Eq.~\eqref{relativeT} constitutes the leading term of an expansion in the operator $A^{-1}$. This constitutes the approximation used in the present work. All throughout the remainder of the paper, $T_{\text{kin}}$ actually stands for $T^{\text{1B}}_{\text{rel}}$ while $V$ really denotes $V+T^{\text{2B}}_{\text{rel}}$.

\subsection{Bogoliubov algebra}

The unitary Bogoliubov transformation connects single-particle $\{c_{p};c^{\dagger}_{p}\}$ to quasiparticle $\{\beta_{\alpha};\beta^{\dagger}_{\alpha}\}$ creation and annihilation operators according to~\cite{ring80a}

\begin{subequations}
\label{e:p2qp}
\begin{align}
\beta_{\alpha}^{\dagger} &= \sum_{p} U_{p\alpha} \, c^{\dagger}_{p} 
 + V_{p\alpha} \,  c_{p} \, , \\
\beta_{\alpha} &= \sum_{p} U^{*}_{p\alpha} \, c_{p} 
 + V^{*}_{p\alpha} \,  c^{\dagger}_{p} \, .
\end{align}
\end{subequations}
Quasiparticle operators obey anticommutation rules such that 
$\{\beta_\alpha,\beta_\beta\}=\{\beta^{\dagger}_{\alpha},\beta^{\dagger}_{\beta}\}=0$ 
and $\{\beta_\alpha,\beta^{\dagger}_\beta\} = \delta_{\alpha\beta}$. 

The Bogoliubov product state, which carries even number-parity as a quantum number, is defined as 
\begin{equation}
\label{e:bogvac}
| \Phi \rangle \equiv \mathcal{C} \displaystyle \prod_{\alpha} \beta_{\alpha} | 0 \rangle,
\end{equation}
and is the vacuum of the quasiparticle operators, i.e. $\beta_\alpha | \Phi \rangle =0$ for all $\alpha$. In Eq.~\eqref{e:bogvac}, $\mathcal{C}$ is a complex normalization. As quasiparticle operators mix particle creation and annihilation operators (see Eq.~\eqref{e:p2qp}), the Bogoliubov vacuum breaks  $U(1)$ symmetry associated with particle conservation, i.e. $| \Phi \rangle$ is not an eigenstate of the particle-number operator, except in the particular limit where it reduces to a Slater determinant.

\subsection{Normal ordering}

A Lagrange term is required to constrain the particle number to the correct value on average, such that the grand canonical potential $\Omega \equiv H - \lambda A$ is used in place of $H$. BCC theory is best formulated in the quasiparticle basis introduced in Eq.~\eqref{e:p2qp} by normal ordering $\Omega$ with respect to $| \Phi \rangle$ via Wick's theorem. Normal ordering an operator with respect to a particle-number-breaking product state invokes two types of elementary contractions, i.e. respectively the normal and anomalous one-body density matrices~\cite{ring80a}
\begin{subequations}
\begin{align}
\label{e:rhowick}
\rho_{qp} \equiv & \frac{\langle \Phi | c^{\dagger}_{p}c_{q} | \Phi \rangle}{\langle \Phi | \Phi \rangle} \, , \\
\label{e:kapwick}
\kappa_{qp} \equiv & \frac{\langle \Phi | c_p c_q | \Phi \rangle}{\langle \Phi | \Phi \rangle} \, .
\end{align}
\end{subequations}
The normal density matrix is hermitian ($\rho^{\dagger} = \rho$) while 
the anomalous density matrix or pairing tensor is skew-symmetric ($\kappa^{T} = - \kappa$).
With recourse to Eq. \eqref{e:p2qp}, these quantities can be written as
\begin{equation}
\label{e:rhokap}
 \rho = V^{*}V^{T}, \: \: \: \: \:  \kappa = V^{*}U^{T}.
\end{equation} 

Once the reference vacuum (i.e. $U$ and $V$ matrices) is specified, the matrix elements of the various normal-ordered contributions to $\Omega$ can be calculated and stored. In the BCC method developed in Sec.~\ref{s:bcc}, the normal-ordered form of $\Omega$ is determined once during the initialization of the calculation, and is then employed consistently throughout the iterative process of solving the BCC equations.  Performing the normal ordering is tedious but straightforward. Explicit expressions of normal-ordered Hamiltonian with respect to a Bogoliubov vacuum have been given in, e.g., \textcite{ring80a}. In the present work, we extend this result in two respects. First, we provide full-fledged expressions for a Hamiltonian containing three-nucleon forces. Second, we express the normal-ordered grand canonical potential in terms of fully antisymmetric matrix elements. The net result, expressed in terms of the fully antisymmetric matrix elements defined in Appendix \ref{s:hamas}, reads as

\begin{widetext}
\begin{subequations}
\label{e:h3qpas}
\begin{align}
\label{e:h3qpasa}
\Omega &\equiv \Omega^{[0]} + \Omega^{[2]} + \Omega^{[4]} + \Omega^{[6]} \\
\nonumber
&\equiv \Omega^{00} + \Big[\Omega^{11} + \{\Omega^{20} + \Omega^{02}\}\Big] \\
\nonumber   &\: \: \: \: \: \: \: \: \: \: \: \: \: + \Big[\Omega^{22} + \{\Omega^{31} + \Omega^{13}\} + 
\{\Omega^{40} + \Omega^{04}\}\Big] \\
   &\: \: \: \: \: \: \: \: \: \: \: \: \: + \Big[\Omega^{33} + \{\Omega^{42} + \Omega^{24}\} + \{\Omega^{51} + \Omega^{15}\} + \{\Omega^{60} + \Omega^{06}\}\Big] \\
&= \Omega^{00} \\
& \: \: \: \: \: \: \: \: \: \: \: \: \: + \frac{1}{1!}\displaystyle\sum_{k_1 k_2} \Omega^{11}_{k_1 k_2}\beta^{\dagger}_{k_1} \beta_{k_2} \\
& \: \: \: \: \: \: \: \: \: \: \: \: \: + \frac{1}{2!}\displaystyle\sum_{k_1 k_2} \Big \{\Omega^{20}_{k_1 k_2} \beta^{\dagger}_{k_1}
 \beta^{\dagger}_{k_2} + \Omega^{02}_{k_1 k_2}   \beta_{k_2} \beta_{k_1} \Big \} \\
& \: \: \: \: \: \: \: \: \: \: \: \: \: + \frac{1}{(2!)^{2}} \displaystyle\sum_{k_1 k_2 k_3 k_4} \Omega^{22}_{k_1 k_2 k_3 k_4} 
   \beta^{\dagger}_{k_1} \beta^{\dagger}_{k_2} \beta_{k_4}\beta_{k_3} \\
   & \: \: \: \: \: \: \: \: \: \: \: \: \: + \frac{1}{3!}\displaystyle\sum_{k_1 k_2 k_3 k_4}\Big \{ \Omega^{31}_{k_1 k_2 k_3 k_4}
   \beta^{\dagger}_{k_1}\beta^{\dagger}_{k_2}\beta^{\dagger}_{k_3}\beta_{k_4} +
   \Omega^{13}_{k_1 k_2 k_3 k_4} \beta^{\dagger}_{k_1} \beta_{k_4} \beta_{k_3} \beta_{k_2}  \Big \} \\
  & \: \: \: \: \: \: \: \: \: \: \: \: \: +  \frac{1}{4!} \displaystyle\sum_{k_1 k_2 k_3 k_4}\Big \{ \Omega^{40}_{k_1 k_2 k_3 k_4}
   \beta^{\dagger}_{k_1}\beta^{\dagger}_{k_2}\beta^{\dagger}_{k_3}\beta^{\dagger}_{k_4}  + 
   \Omega^{04}_{k_1 k_2 k_3 k_4}  \beta_{k_4} \beta_{k_3} \beta_{k_2} \beta_{k_1}  \Big \}  \\
\label{e:h3qpasi}
   & \: \: \: \: \: \: \: \: \: \: \: \: \: + \frac{1}{(3!)^2} \displaystyle\sum_{k_1 k_2 k_3 k_4 k_5 k_6} 
   \Omega^{33}_{k_1 k_2 k_3 k_4 k_5 k_6}
   \beta^{\dagger}_{k_1}\beta^{\dagger}_{k_2}\beta^{\dagger}_{k_3}\beta_{k_6}\beta_{k_5}\beta_{k_4} \\
   & \: \: \: \: \: \: \: \: \: \: \: \: \: + \frac{1}{(2!) (4!)} \displaystyle\sum_{k_1 k_2 k_3 k_4 k_5 k_6} \Big \{ 
   \Omega^{42}_{k_1 k_2 k_3 k_4 k_5 k_6}\beta^{\dagger}_{k_1}\beta^{\dagger}_{k_2}\beta^{\dagger}_{k_3}
   \beta^{\dagger}_{k_4}\beta_{k_6}\beta_{k_5} +  \Omega^{24}_{k_1 k_2 k_3 k_4 k_5 k_6}
   \beta^{\dagger}_{k_1}\beta^{\dagger}_{k_2}\beta_{k_6}\beta_{k_5}\beta_{k_4}\beta_{k_3} \Big \} \\
   & \: \: \: \: \: \: \: \: \: \: \: \: \: + \frac{1}{5!} \displaystyle\sum_{k_1 k_2 k_3 k_4 k_5 k_6} \Big \{ 
   \Omega^{51}_{k_1 k_2 k_3 k_4 k_5 k_6}\beta^{\dagger}_{k_1}\beta^{\dagger}_{k_2}\beta^{\dagger}_{k_3}
   \beta^{\dagger}_{k_4}\beta^{\dagger}_{k_5}\beta_{k_6} +  \Omega^{15}_{k_1 k_2 k_3 k_4 k_5 k_6}
   \beta^{\dagger}_{k_1}\beta_{k_6}\beta_{k_5}\beta_{k_4}\beta_{k_3}\beta_{k_2} \Big \} \\
   & \: \: \: \: \: \: \: \: \: \: \: \: \: + \frac{1}{6!} \displaystyle\sum_{k_1 k_2 k_3 k_4 k_5 k_6} \Big \{ 
   \Omega^{60}_{k_1 k_2 k_3 k_4 k_5 k_6}\beta^{\dagger}_{k_1}\beta^{\dagger}_{k_2}\beta^{\dagger}_{k_3}
   \beta^{\dagger}_{k_4}\beta^{\dagger}_{k_5}\beta^{\dagger}_{k_6} +  \Omega^{06}_{k_1 k_2 k_3 k_4 k_5 k_6}
   \beta_{k_6}\beta_{k_5}\beta_{k_4}\beta_{k_3}\beta_{k_2}\beta_{k_1} \Big \} \, .
   \end{align}
   \end{subequations}
\end{widetext}
Let us now make a set of observations to clarify the content of Eq.~\eqref{e:h3qpas}.
\begin{enumerate}
\item Each term $\Omega^{ij}$ in Eq.~\eqref{e:h3qpas} is characterized by its number $i$ ($j$) of quasiparticle creation (annihilation) operators. Because $\Omega$ has been normal-ordered  with respect to $| \Phi \rangle$, all quasiparticle creation operators (if any) are located to the left of all quasiparticle annihilation operators (if any).  The class $\Omega^{[k]}$ groups all the terms $\Omega^{ij}$ for which $i+j=k$. The first contribution 
\begin{equation}
\Omega^{[0]} = \Omega^{00} = \frac{\langle \Phi | \Omega | \Phi \rangle}{\langle \Phi | \Phi \rangle} 
\end{equation}
denotes the fully contracted part of $\Omega$ and is nothing but a (real) number. 
\item The subscripts of the matrix elements are ordered sequentially, independently of 
the creation or annihilation character of the operators the indices refer to. While quasiparticle creation operators themselves also follow 
sequential order, quasiparticle annihilation operators follow inverse sequential order. In Eq. \eqref{e:h3qpasi}, for example, the three 
creation operators are ordered $\beta^{\dagger}_{k_1}\beta^{\dagger}_{k_2}\beta^{\dagger}_{k_3}$ while 
the three annihilation operators are ordered $\beta_{k_6}\beta_{k_5}\beta_{k_4}$.
\item Matrix elements are fully antisymmetric, i.e.
\begin{eqnarray}
\Omega^{ij}_{k_1 \ldots k_{i} k_{i+1} \ldots k_{i+j}} &=& (-1)^{\sigma(P)}
\Omega^{ij}_{P(k_1 \ldots k_i | k_{i+1} \ldots k_{i+j})} 
\end{eqnarray}
where $\sigma(P)$ refers to the signature of the 
permutation $P$.  The notation $P(\ldots | \ldots)$ denotes a 
separation into the $i$ quasiparticle creation operators and the $j$ quasiparticle annihilation operators such that
permutations are only considered between members of the same group. 
\item Recent ab initio calculations of mid-mass nuclei have made clear that contributions from the three-nucleon interaction need to be included~\cite{Hagen:2007ew,Binder:2013xaa,Hergert:2013uja,Cipollone:2013zma,Soma:2013xha}. Still, computational requirements make it challenging to include them in full. As a result, the typical procedure consists of truncating the normal-ordered Hamiltonian by excluding $\Omega^{[6]}$ such that the dominant effect of the three-nucleon interaction is taken into account through its contribution to $\Omega^{[k]}$ with $k \le 4$.\footnote{While this is strictly true in CC calculations~\cite{Binder:2013oea}, MR-IMSRG calculations of open-shell nuclei as well as SCGF calculations of closed- and open-shell nuclei truncate the Hamiltonian after normal ordering it with respect to a partially~\cite{Hergert:2013uja} or fully correlated state~\cite{Carbone:2013eqa}, respectively.} This is shown to work well in mid-mass closed-shell nuclei, although the omitted part of the three-nucleon interaction may contribute on the same level as the triple corrections~\cite{Binder:2012mk}. Following this procedure, the explicit expressions of $\Omega^{ij}_{k_1 \ldots k_i k_{i+1} \ldots k_{i+j}}$ in terms of interaction and ($U,V$) matrix elements are provided in Appendix~\ref{s:hamas} for $i+j \le 4$. The remaining terms have been derived and can be used eventually to include the residual, i.e. $\Omega^{[6]}$, part of the three-nucleon force.
\end{enumerate}

\subsection{Hartree-Fock-Bogoliubov reference state}

\subsubsection{Variational problem}

The expressions thus far have been formulated for an arbitrary Bogoliubov vacuum (Eq. \eqref{e:bogvac}). In practical applications, one must specify the way this vacuum $| \Phi \rangle$ is actually determined. Several choices are possible:  a Brueckner reference state which maximizes the overlap with the true ground state~\cite{handy89a}, a simple BCS state, or the solution of the variational problem, i.e. using  the Bogoliubov vacuum that solves self-consistent Hartree-Fock-Bogoliubov (HFB) equations~\cite{ring80a} under a set of symmetry requirements. We focus here on the third option. 

The HFB eigenvalue equation can be expressed~\cite{ring80a}
\begin{equation}
\label{e:hfb}
\begin{pmatrix}
h & \Delta \\
-\Delta^{*} & -h^{*}
\end{pmatrix}
\begin{pmatrix}
U_\alpha \\
V_\alpha
\end{pmatrix}
= E_\alpha
\begin{pmatrix}
U_\alpha \\
V_\alpha
\end{pmatrix},
\end{equation}
where columns $(U_\alpha, V_\alpha)$ of the $U$ and $V$ matrices determine the quasiparticle operator $\beta^{\dagger}_\alpha$ 
of Eq. \eqref{e:p2qp}, and where $h$ and $\Delta$ are defined in Eq.~\eqref{variousdefinitions}.  In actual BCC applications, the HFB solution will be utilized as the reference Bogoliubov vacuum.  
Throughout this work, however, a general Bogoliubov vacuum is used to derive BCC equations. Any result depending specifically on the use of a HFB reference state will have the Bogoliubov vacuum denoted as $| \Phi_{\text{HFB}} \rangle$.

\subsubsection{Spectroscopic factors}

Although Bogoliubov states do not carry a definite particle number, it is still useful to 
discuss the spectroscopic content associated with $| \Phi \rangle$.  The spectroscopic factors for the addition (removal) 
of a nucleon are denoted by $\mathfrak{F}^+_\alpha (\mathfrak{F}^-_\alpha)$ and give~\cite{Rotival:2007hp}
\begin{subequations}
\label{e:gensf}
\begin{align}
\mathfrak{F}^+_\alpha &\equiv \displaystyle\sum_p \langle \Phi | c_p | \Phi^\alpha \rangle \langle \Phi^\alpha | c^{\dagger}_p | \Phi 
\rangle = \displaystyle\sum_p \left| U_{p\alpha} \right|^2, \\
\mathfrak{F}^-_\alpha &\equiv \displaystyle\sum_p \langle \Phi | c^{\dagger}_p | \Phi^\alpha \rangle \langle \Phi^\alpha | c_p | \Phi 
\rangle = \displaystyle\sum_p \left| V_{p\alpha} \right|^2,
\end{align}
\end{subequations}
where the odd number-parity states  $| \Phi^\alpha \rangle$ describe the $\text{A} \pm 1$ systems.
 
 \subsubsection{Binding energy}
 \label{s:genhfbbe}
 
The expression of the HFB total energy ${\cal E}_0$ is obtained through the normal ordering of $\Omega$ with respect to $| \Phi_{\text{HFB}} \rangle$ given that $\Omega^{00}= {\cal E}_0 - \lambda \text{A}$ (Eq. \eqref{e:me3defasa}).  The energy can also be computed from the Galitskii-Koltun sum rule at play in self-consistent Gorkov-Green's function theory~\cite{soma11a}.  This alternative formulation provides a check for consistency  and convergence in the solution of the HFB equations, and can be written under the form of a trace over the one-body Hilbert space ${\cal H}_{1}$, i.e.
\begin{eqnarray}
\Omega^{00} &=&  +\frac{1}{4\pi i} \int_C d \omega \; \text{Tr}_{{\cal H}_{1}} \big\{G^{11(0)}(\omega) 
\big[T - \lambda  + \omega \big]\big\} \nonumber \\
&& - \frac{1}{6}\text{Tr}_{{\cal H}_{1}} \big\{\Gamma^{3N} \rho + \Delta^{3N} \kappa^{*}\big\} \, , \label{koltun}
\end{eqnarray}
where $G^{11(0)}(\omega)$ denotes the HFB approximation to the normal Gorkov propagator, while the second line represents the explicit correction to the standard Galitskii-Koltun sum rule due to the presence of three-nucleon forces~\cite{Carbone:2013eqa}. The explicit expressions of the Hartree-Fock $\Gamma^{3N}$ and Bogoliubov $\Delta^{3N}$ fields associated with the three-nucleon force contribution are provided in App.~\ref{s:hamas}. Writing $G^{11(0)}(\omega)$ in its Lehmann representation
\begin{equation}
G^{11(0)}_{ab}(\omega) = \sum_\alpha \frac{U_{a\alpha}U^*_{b\alpha}}{\omega-E_\alpha+i\eta} + 
\frac{V_{a\alpha}^*V_{b\alpha}}{\omega+E_\alpha-i\eta},
\end{equation}
where $\eta$ is an infinitesimally small parameter, the contour integral in Eq.~\eqref{koltun} is effected over the upper-half plane to obtain
\begin{subequations}
\label{e:genksr}
\begin{eqnarray}
{\cal E}_{0} &=&  + \frac{1}{2} \Big[\sum_{pq} t_{pq} \, \rho_{qp}  - \sum_{\alpha} \big(E_\alpha - \lambda\big) \, \mathfrak{F}^-_\alpha \Big] \nonumber \\
&& - \frac{1}{6} \Big[\sum_{pq}  \Gamma^{3N}_{pq} \, \rho_{qp} + \Delta^{3N}_{pq} \, \kappa^{*}_{qp}\Big] \, .
\end{eqnarray}
\end{subequations}

\section{Coupled cluster theory}
\label{s:bcc}

\subsection{Coupled cluster ansatz}

In standard coupled cluster (CC) theory, the ground-state wavefunction of the system is written 
in the exponentiated form
\begin{equation}
| \Psi \rangle \equiv e^{T} | \Phi \rangle \, ,
\end{equation}
where $| \Phi \rangle$ is a Slater determinant and where the cluster operator $T \equiv T_1 + T_2 + T_3 + \ldots$ is the sum of connected n-tuple excitation operators of the form~\cite{shavitt09a} 
\begin{subequations}
\label{e:cct}
\begin{align}
\label{e:ccta}
T_1 &\equiv \frac{1}{(1!)^{2}}\sum_{ia}t^{a}_{i} c^{\dagger}_a c_i \, , \\
T_2 &\equiv \frac{1}{(2!)^{2}}\displaystyle\sum_{ijab}t^{ab}_{ij} c^{\dagger}_a c_i c^{\dagger}_b c_j \, , \\
T_3 &\equiv \frac{1}{(3!)^{2}}\displaystyle\sum_{ijkabc} t^{abc}_{ijk} c^{\dagger}_a c_i c^{\dagger}_b c_j c^{\dagger}_c c_k \, ,
\end{align}
\end{subequations}
etc., where the amplitudes $t^{ab\ldots}_{ij\ldots}$ are the unknowns to be determined. As $T$ is expressed in normal-ordered form with respect to the Slater determinant $| \Phi \rangle$, intermediate normalization $\langle \Phi | \Psi \rangle=1$ is in order. Occupied (hole) and unoccupied (particle) single-particle states of $| \Phi \rangle$ can be distinguished; i.e. label indices $a,b,c\ldots$ specifically denote particle states while labels $i,j,k\ldots$ refer to hole states. As was already clear from above, the notation $p,q,r\ldots$ is used when referring to a general set of single-particle basis states.

This traditional CC scheme is presently extended to a Bogoliubov setting where the ground-state wavefunction of the system is written in the form
\begin{equation}
\label{e:bccwf}
| \Psi \rangle \equiv e^{\mathcal{T}} | \Phi \rangle \, ,
\end{equation}
where $ | \Phi \rangle$ denotes now the Bogoliubov vacuum of Eq. \eqref{e:bogvac}, and where the {\it quasiparticle} cluster 
operator $\mathcal{T} \equiv \mathcal{T}_1 +\mathcal{T}_2 + \mathcal{T}_3 + \ldots$ is defined by
\begin{subequations}
\label{e:qptv1as}
\begin{align}
\label{e:qptv1asa}
\mathcal{T}_1 &\equiv \frac{1}{2!}\displaystyle\sum_{k_1 k_2}t_{k_1 k_2} 
\beta^{\dagger}_{k_1}\beta^{\dagger}_{k_2} \, , \\
\mathcal{T}_2 &\equiv \frac{1}{4!}\displaystyle\sum_{k_1 k_2 k_3 k_4}t_{k_1 k_2 k_3 k_4} 
\beta^{\dagger}_{k_1} \beta^{\dagger}_{k_2}\beta^{\dagger}_{k_3} \beta^{\dagger}_{k_4} \, , \\
\mathcal{T}_3 &\equiv \frac{1}{6!}\displaystyle\sum_{k_1 k_2 k_3 k_4 k_5 k_6} 
t_{k_1 k_2 k_3 k_4 k_5 k_6} \beta^{\dagger}_{k_1} \beta^{\dagger}_{k_2} 
\beta^{\dagger}_{k_3}\beta^{\dagger}_{k_4}\beta^{\dagger}_{k_5}\beta^{\dagger}_{k_6} \, ,
\end{align}
\end{subequations}
etc.  The quasiparticle amplitudes $t_{k_1 k_2 \ldots}$, which need to be determined, 
are fully antisymmetric, i.e. $t_{k_1 k_2 \ldots} = 
(-1)^{\sigma(P)} t_{P(k_1 k_2 \ldots)}$, resulting in the $(2n!)^{-1}$ normalization factor in the definition of $\mathcal{T}_n$. Similarly to standard CC theory, the operator $\mathcal{T}_n$ is in normal-ordered form with respect to the Bogoliubov vacuum $| \Phi \rangle$, which leads to intermediate normalization $\langle \Phi | \Psi \rangle=1$.

\subsection{Similarity-transformed Hamiltonian}

Given the BCC ansatz of Eq. \eqref{e:bccwf}, the Schr\"odinger equation $\Omega | \Psi \rangle = \Omega_0 | \Psi \rangle$ can be written as
\begin{equation}
\label{e:eiggcp}
\Omega \, e^{\mathcal{T}} | \Phi \rangle = \Omega_0 \, e^{\mathcal{T}} | \Phi \rangle \, .
\end{equation}
Operating from the left with $e^{-\mathcal{T}}$ results in 
\begin{equation}
\label{e:schreff}
e^{- \mathcal{T}} \Omega \, e^{\mathcal{T}} | \Phi \rangle = \Omega_0 | \Phi \rangle \, .
\end{equation}
an eigenvalue equation for the non-hermitian similarity-transformed grand canonical 
potential 
\begin{equation}
\label{e:bcceff}
\bar{\Omega} \equiv e^{- \mathcal{T}} \Omega \, e^{\mathcal{T}}
\end{equation}
with ground-state eigenvalue $\Omega_0$ and right-eigenfunction $| \Phi \rangle$.  This operator is referred to as the BCC effective grand potential.  

The Baker-Campbell-Hausdorff expansion allows one to write
\begin{eqnarray}
\label{e:hambched}
\bar{\Omega} &=& \Omega + [\Omega,\mathcal{T}] + \frac{1}{2!}[[\Omega,\mathcal{T}],\mathcal{T}]   \\ 
&&\hspace{0.35cm} + 
\frac{1}{3!}[[[\Omega,\mathcal{T}],\mathcal{T}],\mathcal{T}] + 
\frac{1}{4!}[[[[\Omega,\mathcal{T}],\mathcal{T}],\mathcal{T}],\mathcal{T}] + \ldots, \nonumber
\end{eqnarray}
which is an infinite sum of nested commutators. Applying Wick's theorem, and given that $\mathcal{T}_m$ consists only of quasiparticle creation operators  such that $[\mathcal{T}_m,\mathcal{T}_n]=0$ for all $m,n$, only terms consisting of at least one contraction between $\Omega$ and each $\mathcal{T}$ operator in the nested commutators remain. This results in a natural termination  of the infinite expansion in Eq. \eqref{e:hambched}.  The grand canonical potential being presently limited to $\Omega^{[k]}$ with $k = 0,2,4,$ Eq. \eqref{e:bcceff} terminates exactly after the term containing four nested commutators.\footnote{If $\Omega^{[6]}$ were to be included, the truncation of the expansion would still occur, but terms with as many as six nested commutators would contribute.} Because non-zero contractions require quasiparticle operators in the  form $\langle \Phi | \beta_{k_1} \beta^{\dagger}_{k_2} | \Phi \rangle$, surviving terms necessarily contain $\Omega$ as the leftmost operator.  Thus, Eq. \eqref{e:bcceff} can be rewritten
\begin{subequations}
\begin{align}
\bar{\Omega} &= \Omega + \Big(\Omega \mathcal{T}\Big)_{\text{C}} + \frac{1}{2!}\Big(\Omega\mathcal{TT}\Big)_{\text{C}} \nonumber \\
& \hspace{0.75cm} + \frac{1}{3!} \Big(\Omega\mathcal{TTT}\Big)_{\text{C}} + \frac{1}{4!}\Big(\Omega \mathcal{TTTT}\Big)_{\text{C}}  \, ,
\end{align}
\end{subequations}
such that $\bar{\Omega} \equiv (\Omega e^{\mathcal{T}})_{\text{C}}$, where the subscript C denotes that only connected terms eventually contribute, i.e. $\Omega$ must have at least one contraction with each $\mathcal{T}$ operator.

\subsection{Bogoliubov coupled cluster equations}
\label{s:bcceqs}

Operating on Eq. \eqref{e:schreff} from the left with $\langle \Phi |$ and $\langle \Phi^{\alpha \beta \ldots} |$ produces the BCC energy equation 
\begin{equation}
\label{e:genbcce}
\langle \Phi | \bar{\Omega}_N  | \Phi \rangle _{\text{C}} = \Delta \Omega_0 
\end{equation} 
along with equations to determine the n-tuple amplitudes
\begin{equation}
\label{e:genbccamp}
\langle \Phi^{\alpha \beta \ldots} | \bar{\Omega}_N | \Phi \rangle _{\text{C}} = 0 \, ,
\end{equation} 
respectively, where
\begin{equation}
\label{e:qpexbv}
| \Phi^{\alpha \beta \ldots} \rangle \equiv \beta^{\dagger}_{\alpha} \beta^{\dagger}_{\beta} \ldots | \Phi \rangle \, .
\end{equation} 
In Eqs.~\eqref{e:genbcce} and~\eqref{e:genbccamp},  one works with
\begin{subequations}
\begin{align}
\bar{\Omega}_N &\equiv e^{\mathcal{-T}}(\Omega - \Omega^{00})e^{\mathcal{T}} \\
&\equiv e^{\mathcal{-T}} \Omega_N e^{\mathcal{T}} \\
&\equiv \big(\Omega_N e^{\mathcal{T}} \big)_{\text{C}} \, ,
\end{align}
\end{subequations}
which eliminates the unnecessary evaluation of terms involving the trivial contribution $\Omega^{00}$ to the normal-ordered grand canonical potential.  The total ground-state energy $E_0$ is eventually obtained from 
\begin{subequations}
\label{e:e0def}
\begin{eqnarray}
\Omega_0 &=& \Omega^{00} + \Delta \Omega_0 \, , \\
&\equiv& E_0 - \lambda \text{A} \, .
\end{eqnarray}
\end{subequations}
It is important to note here, as will be discussed below, that the chemical potential obtained from the solution of the HFB equations is in principle different from that obtained in the solution of the BCC equations.  Therefore, one must be careful in evaluating Eq.~\eqref{e:e0def} to obtain $E_0$ correctly, 
which is of course independent of the chemical potential and therefore can be written $E_0 = {\cal E}_{0} + \Delta H_0$, with $\Delta H_0$ obtained analogously to $\Delta \Omega_0$ from\footnote{In practice, it is more straightforward to determine $E_0$ directly from Eq. \eqref{e:e0def}, evaluating
$\Omega_0$ at the BCC chemical potential.}
\begin{equation}
\label{e:genh0}
\langle \Phi | \bar{H}_N  | \Phi \rangle _{\text{C}} = \Delta H_0 \, .
\end{equation}

\subsection{Constraint on particle number}
\label{Nconstrain}

The energy and amplitude equations (Eqs. \eqref{e:genbcce} and \eqref{e:genbccamp}) must be solved under the constraint $\langle \Psi | A | \Psi \rangle / \langle \Psi | \Psi \rangle = \text{A}$. Even when this condition is imposed on the reference state $| \Phi \rangle$, it is not automatically maintained for the coupled cluster wavefunction, which must thus be constrained as well. In practice, of course, this must be done separately for both the neutron number $\text{N}$ and the proton number $\text{Z}$. In this formal presentation, $\text{A}$ stands for either of them.

It is thus mandatory to compute the average value of the one-body operator $A$ repeatedly while finding the cluster amplitudes iteratively, and this for any truncation scheme of interest (see below). There are various ways of attacking this problem. One possibility relies on the Hellmann-Feynman theorem that accesses the average value of $A$ via the numerical derivative of $\Omega_0$ with respect to the chemical potential. However, the Hellman-Feynman theorem can exhibit instabilities near phase transitions, such as those employed by our spontaneous breaking of $U(1)$ symmetry. In the optimal procedure \cite{shavitt09a}, 
\begin{eqnarray}
\label{e:leftschreff}
\text{A} &=& \frac{\langle \Psi | A | \Psi \rangle}{\langle \Psi | \Psi \rangle} = \frac{\langle \Phi | e^{\mathcal{T}^{\dagger}} A e^{\mathcal{T}} | \Phi \rangle}{\langle \Phi | e^{\mathcal{T}^{\dagger}} e^{\mathcal{T}} | \Phi \rangle} \\
&=& \frac{\langle \Phi | A | \Phi \rangle}{\langle \Phi | \Phi \rangle} + \langle \Phi | e^{\mathcal{T}^{\dagger}} A_N e^{\mathcal{T}} | \Phi \rangle_{\text{C}} \\
&=& \frac{\langle \Phi | A | \Phi \rangle}{\langle \Phi | \Phi \rangle} + \langle \Phi |(1+ \Lambda) e^{\mathcal{-T}} A_N e^{\mathcal{T}} | \Phi \rangle_{\text{C}} \, ,
\end{eqnarray}
where the de-excitation operator $\Lambda = \Lambda_1 + \Lambda_2 + \ldots$ is determined from the solution of the eigenvalue problem 
for the left ground state of $\bar{\Omega}$~\cite{shavitt09a}, and where the normal-ordered part of any operator $O_N = O - \langle \Phi | O | \Phi \rangle$.
We will describe the evaluation of the particle number in this approach, but 
eventually use an approximation to evaluate the left ground state.

We first normal order the particle-number operator with respect to $| \Phi \rangle$, i.e.
\begin{subequations}
\label{normalorderedA}
\begin{eqnarray}
A &\equiv& A^{[0]} + A^{[2]}  \\
&\equiv& A^{00} \\
&& + \frac{1}{1!} \sum_{k_1 k_2} A^{11}_{k_1 k_2}\beta^{\dagger}_{k_1} \beta_{k_2} \\
&& + \frac{1}{2!}\sum_{k_1 k_2} \Big \{A^{20}_{k_1 k_2} \beta^{\dagger}_{k_1}
 \beta^{\dagger}_{k_2} + A^{02}_{k_1 k_2}   \beta_{k_2} \beta_{k_1} \Big \} \, ,
\end{eqnarray}
\end{subequations}
where the expression of the matrix elements are provided in App.~\ref{1Boperator}. Given that the reference contribution is $\langle \Phi | A | \Phi \rangle/\langle \Phi | \Phi \rangle =  A^{00} = \text{Tr}\big[ \rho\big]$, the correction to it is
\begin{eqnarray}
\delta \text{A} &=&  \langle \Phi | (1+\Lambda) \, e^{- \mathcal{T}} A^{[2]} e^{\mathcal{T}} | \Phi \rangle_{\text{C}} \, . \label{deltaA}
\end{eqnarray}

If the Bogoliubov vacuum satisfies the correct particle number on average, the correction $\delta \text{A}$ must be constrained to zero.   In practice, lower energies are obtained using this method (in the approximation discussed in Sec. \ref{BCCSDsection}) relative to those obtained when solving the BCC system of equations via the Hellman-Feynman theorem to evaluate the particle number.  This emphasizes the danger in employing the Hellmann-Feynman theorem near phase transitions.

In addition to constraining the average particle number, it is of interest to monitor the breaking of the symmetry by computing the variance associated with the operator $A$. In the same spirit, our solution will be allowed to break good angular momentum such that it is of interest to monitor the average value of the operator $J^2$, which informs us directly on the breaking of rotational symmetry when targeting the $J^{\pi}=0^+$ ground state of an even-even nucleus.  
From the operators $A$ and $A^2$, the particle-number variance $\Delta \text{A}^2$ is obtained via
\begin{eqnarray}
\Delta \text{A}^2 &=& \frac{\langle \Psi | A^2 | \Psi \rangle}{\langle \Psi | \Psi \rangle} - \left(\frac{\langle \Psi | A | \Psi \rangle}{\langle \Psi | \Psi \rangle}\right)^2 \, .
\label{e:vareq}
\end{eqnarray}

\subsection{Computing observables}
\label{opgen}

While in principle the expectation value of any operator can be expressed in terms of density matrices and normal-ordered matrix elements of the operator, one can instead evaluate the expectation value by exploiting the BCC energy and amplitude equations (Eqs.~\eqref{e:genbcce} and~\eqref{e:genbccamp}).  

We want to evaluate the expectation value 
\begin{eqnarray}
\text{O} = \frac{ \langle \Psi | O | \Psi \rangle}{\langle \Psi | \Psi \rangle} \, ,
\end{eqnarray}
which can be written \cite{shavitt09a}
\begin{subequations}
\label{e:splitoop}
\begin{eqnarray}
\text{O} &=& \langle \Phi | O | \Phi \rangle + \langle \Phi | (1+\Lambda) \, e^{- \mathcal{T}} O_N e^{\mathcal{T}} | \Phi \rangle_{\text{C}} \\
             &=& \text{O}_{\text{ref}} + \Delta \text{O} \, ,
\end{eqnarray}
\end{subequations}
where at this point the operator $O$ is completely general.  The reference contribution $\text{O}_{\text{ref}}$ can be evaluated straightforwardly. To 
evaluate the second term on the righthand side of Eq. \eqref{e:splitoop}, let us first define the Fock-space projection operators 
\begin{subequations}
\label{e:defpq}
\begin{eqnarray}
P &=& | \Phi \rangle \langle \Phi | \\
Q &=& \displaystyle\sum_{\alpha} | \Phi^{\alpha} \rangle \langle \Phi^{\alpha} | + \frac{1}{2!} \sum_{\alpha \beta}
 | \Phi^{\alpha \beta} \rangle \langle \Phi^{\alpha \beta} | \\
 && + \frac{1}{3!} \sum_{\alpha \beta \gamma} 
  | \Phi^{\alpha \beta \gamma} \rangle \langle \Phi^{\alpha \beta \gamma} |  + \frac{1}{4!} \sum_{\alpha \beta \gamma \delta}
 | \Phi^{\alpha \beta \gamma \delta} \rangle \langle \Phi^{\alpha \beta \gamma \delta} | \nonumber \\
&& + \ldots \nonumber \, ,
\end{eqnarray}
\end{subequations}
which satisfy the identity $1=P+Q$.  Inserting this identity into the second term of Eq. \eqref{e:splitoop},
\begin{subequations}
\label{e:genopexp}
\begin{eqnarray}
\Delta \text{O} &=& \langle \Phi | (1+\Lambda) \, [P+Q] \, e^{- \mathcal{T}} O_N e^{\mathcal{T}} | \Phi \rangle_{\text{C}} \\
&=& \langle \Phi | (1+\Lambda) \, | \Phi \rangle \langle \Phi | \, e^{- \mathcal{T}} O_N e^{\mathcal{T}} | \Phi \rangle_{\text{C}} \\
&& + \langle \Phi | (1+\Lambda) \, Q \, e^{- \mathcal{T}} O_N e^{\mathcal{T}} | \Phi \rangle_{\text{C}} \nonumber \\
&=& \langle \Phi | \bar{O}_N | \Phi \rangle_{\text{C}} 
+ \langle \Phi | \Lambda \, Q \, \bar{O}_N | \Phi \rangle_{\text{C}} \nonumber \, .
\end{eqnarray}
\end{subequations}
While we have included terms with both an odd and even number of quasiparticle creation operators in our definition of $Q$ in Eq. \eqref{e:defpq}, the odd terms do not contribute in Eq. \eqref{e:genopexp}, since the 
Bogoliubov reference state carries even number-parity as a quantum number and each component of $\Omega$ and $\mathcal{T}$ conserves number-parity.  Thus, we only access the terms which sum over an even number of 
quasiparticle excitations. 
For the operator $O=\Omega$, one can observe that the BCC equations (Eqs. \eqref{e:genbcce} and \eqref{e:genbccamp}) are reproduced, such that the energy 
from Eq. \eqref{e:e0def} is recovered.  In practice, as will be discussed in Sec. \ref{BCCSDsection}, this form is convenient to obtain the expectation value of 
one- and two-body operators, such as $A$ and $A^2$.

\section{BCC with singles and doubles}
\label{BCCSDsection}

\subsection{Truncation scheme}

Bogoliubov coupled cluster theory is formally exact at this stage.  The approximation in practical calculations results from a truncation of 
the operator $\mathcal{T}$ to a limited number of n-tuple terms $\mathcal{T}_n$.  The simplest approach truncates all terms beyond the one-body operator $\mathcal{T}_1$. In connection with the nomenclature of standard coupled cluster theory, this truncation scheme will be referred to as Bogoliubov coupled cluster with singles (BCCS). The present aim is to implement Bogoliubov coupled cluster with singles and doubles (BCCSD), where $\mathcal{T}^{\text{BCCSD}} = \mathcal{T}_1 + \mathcal{T}_2$. The BCCSD scheme encompasses the most common standard CC approximation, i.e. CCSD, as a particular case. The extension of standard approximations for the treatment of triples, e.g. $\Lambda$-CCSD(T)~\cite{taube08a} or CR-CC(2,3)~\cite{piecuch05a}, in the context of Bogoliubov coupled cluster theory is expected to provide an excellent approximation to open-shell systems. These developments, however, are postponed to future works. 

In the present section, the pedestrian approach to obtaining algebraic forms of BCCSD equations is followed. Eventually, it is inefficient to code the equations in the fully expanded form thus provided such that one relies on the introduction of so-called {\it intermediates}~\cite{shavitt09a}. The latter have the benefit to limit the computational cost and make the equations more compact and readable. The BCCSD equations expressed in terms of intermediates are provided in App.~\ref{quasilinearBCCSD}. It should be further noted that schemes with greater truncation, such as BCCS or Bogoliubov coupled cluster with doubles (BCCD), can be easily deduced from the set of BCCSD equations provided below.

\subsection{Expanded BCCSD equations}

Truncating the cluster operator according to $\mathcal{T}^{\text{BCCSD}}$, the correction to the unperturbed energy (Eq.~\eqref{e:genbcce}) reads
\begin{equation}
\label{e:ebccsd}
\Delta\Omega_0 = 
\langle \Phi | \Omega_N \big(\mathcal{T}_1 + \mathcal{T}_2 + \tfrac{1}{2}\mathcal{T}^2_1\big) | \Phi \rangle _{\text{C}} \, .
\end{equation}
This expression is in fact formally exact even when higher n-tuple cluster operators are included, at least as long as the grand canonical potential is restricted to terms $\Omega^{[k]}$ with $k \le 4$. The inclusion of higher terms in 
$\mathcal{T}$ would affect the energy only indirectly by modifying the quasiparticle amplitudes 
$t_{k_1 k_2}$ and $t_{k_1 k_2 k_3 k_4}$ entering 
Eq. \eqref{e:ebccsd}. Exploiting the full antisymmetry of the quasiparticle amplitudes 
(Eq. \eqref{e:qptv1as}) and of the matrix elements of the grand canonical potential (Appendix \ref{s:hamas}), 
the application of Wick's theorem permits the algebraic expansion of the energy equation (Eq.~\eqref{e:ebccsd}) under the form
\begin{eqnarray}
\label{e:bccsde}
\Delta\Omega_0 &=&  \frac{1}{2} \sum_{k_1 k_2} \Omega^{02}_{k_1 k_2} 
t_{k_1 k_2} \nonumber \\
&& + \frac{1}{4!} \sum_{k_1 k_2 k_3 k_4} \Omega^{04}_{k_1 k_2 k_3 k_4} 
t_{k_1 k_2 k_3 k_4} \nonumber \\
&& + \frac{1}{8} \sum_{k_1 k_2 k_3 k_4} 
\Omega^{04}_{k_1 k_2 k_3 k_4} t_{k_1 k_2} t_{k_3 k_4} \, .
\end{eqnarray}
The singles and doubles quasiparticle amplitudes, respectively $t_{k_1 k_2}$ and $t_{k_1 k_2 k_3 k_4}$, remain to be determined by applying Eq.~\eqref{e:genbccamp} for two ($\langle \Phi^{\alpha \beta}|$) and four ($\langle \Phi^{\alpha \beta \gamma \delta}|$) quasiparticle states. The single-excitation\footnote{To connect with the vocabulary at play in standard CC theory, the equation of motion obtained by left projecting with two (four) quasiparticle states is said to provide the single- (double-) excitation amplitudes.} amplitude equations are given by 
\begin{equation}
\label{e:bccsdt1}
0 = \langle \Phi^{\alpha \beta} | \Omega_N(1+\mathcal{T}_1 + \tfrac{1}{2}\mathcal{T}^2_1 + \tfrac{1}{3!}\mathcal{T}^3_1
+ \mathcal{T}_2 + \mathcal{T}_1\mathcal{T}_2) | \Phi \rangle _{\text{C}},
\end{equation}
while the double-excitation amplitude equations are
\begin{eqnarray}
\label{e:bccsdt2}
0 &=& \langle \Phi^{\alpha \beta \gamma \delta} | \Omega_N(1+\mathcal{T}_1 + \mathcal{T}_2 + 
\tfrac{1}{2}\mathcal{T}^2_1 + \tfrac{1}{2}\mathcal{T}^2_2    \\
&& \hspace {1.6cm} +\mathcal{T}_1\mathcal{T}_2+ \tfrac{1}{3!}\mathcal{T}^3_1 + \tfrac{1}{4!}\mathcal{T}^4_1
+ \tfrac{1}{2} \mathcal{T}^2_1 \mathcal{T}_2 ) | \Phi \rangle _{\text{C}} \, . \nonumber
\end{eqnarray}
Applying Wick's theorem, one obtains the expanded algebraic form of the single-excitation amplitude equations
\newpage
\begin{widetext}
\begin{align}
\label{e:bccsd2qpex}
\nonumber
0 &= \Omega^{20}_{\alpha \beta} \\
\nonumber
 & \hspace{0.3cm}   + \displaystyle\sum_{k_1} \big[\Omega^{11}_{\alpha k_1} 
t_{k_1 \beta} + \Omega^{11}_{\beta k_1} t_{\alpha k_1} \big] \\
\nonumber
 & \hspace{0.3cm}  + 
\frac{1}{2}\sum_{k_1 k_2} \Big[ \Omega^{22}_{\alpha \beta k_1 k_2} t_{k_1 k_2} + 
\Omega^{02}_{k_1 k_2} \big( t_{\alpha \beta k_1 k_2} + 2 t_{\alpha k_1}
t_{k_2 \beta} \big) \Big] \\
\nonumber
 & \hspace{0.3cm} +\frac{1}{6}\displaystyle\sum_{k_1 k_2 k_3} \Big[ \Omega^{13}_{\alpha k_1 k_2 k_3} \big(
 t_{k_1 \beta k_2 k_3} + 3 t_{k_1 \beta}t_{k_2 k_3} \big) +
 \Omega^{13}_{\beta k_1 k_2 k_3} \big(t_{\alpha k_1 k_2 k_3} 
 + 3 t_{\alpha k_1}t_{k_2 k_3} \big)\Big] \\
 & \hspace{0.3cm} + \frac{1}{12} \displaystyle\sum_{k_1 k_2 k_3 k_4} \Omega^{04}_{k_1 k_2 k_3 k_4} \big(
  2t_{\alpha k_1}t_{k_2 \beta k_3 k_4} + 2t_{\beta k_1}
  t_{\alpha k_2 k_3 k_4} + 3 t_{k_1 k_2}t_{\alpha \beta k_3 k_4}
  + 6 t_{\alpha k_1} t_{k_2 k_3} t_{k_4 \beta} \big) \, ,
\end{align}
and of the double-excitation amplitude equations
\begin{align}
\label{e:bccsd4qpex}
\nonumber
0 &= \Omega^{40}_{\alpha \beta \gamma \delta} \\
\nonumber
 & \hspace{0.3cm}  + \displaystyle\sum_{k_1} 
\Big[\Omega^{31}_{\alpha \beta \gamma k_1} t_{k_1 \delta} + 
\Omega^{31}_{\alpha \beta \delta k_1} t_{\gamma k_1} + 
\Omega^{31}_{\alpha \gamma \delta k_1} t_{k_1 \beta} + 
\Omega^{31}_{\beta \gamma \delta k_1} t_{\alpha k_1} \Big] \\
\nonumber
 & \hspace{0.3cm}+ \displaystyle\sum_{k_1} \Big[ \Omega^{11}_{\alpha k_1} t_{k_1 \beta \gamma \delta} +
\Omega^{11}_{\beta k_1} t_{\alpha k_1 \gamma \delta} + 
\Omega^{11}_{\gamma k_1} t_{\alpha \beta k_1 \delta} + 
\Omega^{11}_{\delta k_1} t_{\alpha \beta \gamma k_1} \Big] \\
\nonumber
& \hspace{0.3cm} + \frac{1}{2} \displaystyle\sum_{k_1 k_2} \Big[\Omega^{22}_{\alpha \beta k_1 k_2} 
\big(t_{k_1 k_2 \gamma \delta} + 2t_{\gamma k_1}t_{k_2 \delta} \big) 
+ \Omega^{22}_{\alpha \gamma k_1 k_2} \big(t_{k_1 k_2 \delta \beta} + 
2t_{k_1 \beta}t_{k_2 \delta} \big) + \Omega^{22}_{\alpha \delta k_1 k_2} 
\big(t_{k_1 k_2 \beta \gamma} + 2t_{k_1 \beta}t_{\gamma k_2} \big) \\
\nonumber
& \hspace{1.8cm}+ \Omega^{22}_{\beta \gamma k_1 k_2} \big( t_{k_1 k_2 \alpha \delta} +
2t_{\alpha k_1}t_{k_2 \delta} \big)+
\Omega^{22}_{\beta \delta k_1 k_2} \big(t_{k_1 k_2 \gamma \alpha} + 
2t_{\alpha k_1}t_{\gamma k_2} \big) + 
\Omega^{22}_{\gamma \delta k_1 k_2} \big(t_{k_1 k_2 \alpha \beta}+
2t_{\alpha k_1}t_{k_2 \beta} \big) \Big] \\
\nonumber
& \hspace{0.3cm}+ \displaystyle\sum_{k_1 k_2} \Omega^{02}_{k_1 k_2} \big[ t_{\alpha k_1} 
t_{k_2 \beta \gamma \delta} + t_{\beta k_1} t
_{\alpha k_2 \gamma \delta} + t_{\gamma k_1} t_{\alpha \beta k_2 \delta} +
t_{\delta k_1} t_{\alpha \beta \gamma k_2} \big] \\
\nonumber
 & \hspace{0.3cm}+ \frac{1}{2} \displaystyle\sum_{k_1 k_2 k_3} \big[ \Omega^{13}_{\alpha k_1 k_2 k_3}
 \big( t_{k_1 \beta}t_{k_2 k_3 \gamma \delta} + 
 t_{k_1 \gamma}t_{k_2 \beta k_3 \delta} + 
 t_{k_1 \delta}t_{k_2 \beta \gamma k_3} +
 t_{k_1 k_2}t_{k_3 \beta \gamma \delta} + 
2 t_{k_1 \gamma} t_{k_2 \beta}t_{k_3 \delta} \big) \\
 \nonumber
  & \hspace{2.0cm}+ \Omega^{13}_{\beta k_1 k_2 k_3}
 \big( t_{\alpha k_1}t_{k_2 k_3 \gamma \delta} + 
 t_{k_1 \gamma}t_{\alpha k_2 k_3 \delta} + 
 t_{k_1 \delta}t_{\alpha k_2 \gamma k_3} +
 t_{k_1 k_2}t_{\alpha k_3 \gamma \delta} + 
2 t_{k_1 \alpha} t_{k_2 \gamma}t_{k_3 \delta} \big) \\
 \nonumber
  & \hspace{2.0cm}+ \Omega^{13}_{\gamma k_1 k_2 k_3}
 \big( t_{\alpha k_1}t_{k_2 \beta k_3 \delta} + 
 t_{\beta k_1}t_{\alpha k_2 k_3 \delta} + 
 t_{k_1 \delta}t_{\alpha \beta k_2 k_3} +
 t_{k_1 k_2}t_{\alpha \beta k_3 \delta} + 
 2t_{k_1 \alpha} t_{\beta k_2}t_{k_3 \delta} \big) \\
 \nonumber
 & \hspace{2.0cm}+ \Omega^{13}_{\delta k_1 k_2 k_3}
 \big( t_{\alpha k_1}t_{k_2 \beta \gamma k_3} + 
 t_{\beta k_1}t_{\alpha k_2 \gamma k_3} + 
 t_{\gamma k_1}t_{\alpha \beta k_2 k_3} +
 t_{k_1 k_2}t_{\alpha \beta \gamma k_3} + 
2 t_{k_1 \alpha} t_{k_2 \beta}t_{k_3 \gamma} \big) \Big] \\
   \nonumber
   & \hspace{0.3cm}+ \frac{1}{24} \displaystyle\sum_{k_1 k_2 k_3 k_4} \Omega^{04}_{k_1 k_2 k_3 k_4}
   \Big[ t_{k_4 \beta \gamma \delta} \big(4t_{\alpha k_1 k_2 k_3} +
   12t_{k_1 k_2}t_{\alpha k_3} \big) + 
   t_{\alpha k_4 \gamma \delta} \big(4t_{\beta k_1 k_2 k_3} + 
   12t_{k_1 k_2}t_{\beta k_3} \big) \\
   \nonumber
   & \hspace{3.9cm}+ t_{\alpha \beta k_4 \delta}(4t_{\gamma k_1 k_2 k_3} +
   12t_{k_1 k_2}t_{\gamma k_3} \big) + 
   t_{\alpha \beta \gamma k_4} \big(4t_{\delta k_1 k_2 k_3} + 
   12t_{k_1 k_2}t_{\delta k_3} \big) \\
   \nonumber
   & \hspace{3.9cm}+ t_{\alpha \beta k_3 k_4} \big(3 t_{k_1 k_2 \gamma \delta} +
   12t_{k_1 \gamma}t_{\delta k_2} \big) + 
   t_{\alpha k_3 \gamma k_4} \big(3 t_{k_1 \beta \delta k_2} +
   12t_{k_1 \beta}t_{\delta k_2} \big) \\
   \nonumber
    & \hspace{3.9cm}+ t_{\alpha k_3 k_4 \delta} \big(3 t_{k_1 \beta \gamma k_2} +
   12t_{k_1 \beta}t_{\gamma k_2} \big) +
   t_{k_3 \beta \gamma k_4} \big(3 t_{\alpha k_1 k_2 \delta} +
   12t_{k_1 \alpha}t_{\delta k_2} \big) \\
   \nonumber
   & \hspace{3.9cm}+ t_{k_3 \beta k_4 \delta} \big(3 t_{\alpha k_1 k_2 \gamma} +
   12t_{k_1 \alpha}t_{\gamma k_2} \big) +
   t_{k_3 k_4 \gamma \delta} \big(3 t_{\alpha \beta k_1 k_2} +
   12t_{k_1 \alpha}t_{\beta k_2} \big) \\
   & \hspace{3.9cm}+ 24t_{ k_1 \alpha}t_{k_2 \beta}t_{k_3 \gamma}
   t_{k_4 \delta} \Big] \, .
 \end{align}
\end{widetext}
The solution of these equations, nonlinear in the quasiparticle amplitudes, can be found iteratively to compute the energy. Doing so requires a zeroth iteration, i.e. an initialization of the quasiparticle amplitudes.  Motivated by perturbation theory, the off-diagonal part of $\Omega^{11}$ is neglected in Eqs.~\eqref{e:bccsd2qpex} and~\eqref{e:bccsd4qpex} along with the nonlinear terms, leading to the two initial conditions
\begin{subequations}
\label{e:tinitbccsd}
\begin{align}
\label{e:tinitbccsda}
t_{\alpha \beta} &= - \frac{\Omega^{20}_{\alpha \beta}}{\Omega^{11}_{\alpha \alpha} 
+ \Omega^{11}_{\beta \beta}} \; , \\
\label{e:tinitbccsdb}
t_{\alpha \beta \gamma \delta} &= - \frac{\Omega^{40}_{\alpha \beta \gamma \delta} + 
(1 + P_{\gamma \delta} - P_{\beta \delta} + P_{\alpha \delta}) \Omega^{31}_{\alpha \beta \gamma \delta} t_{\delta \delta}} 
{\Omega^{11}_{\alpha \alpha}  + \Omega^{11}_{\beta \beta}  + 
\Omega^{11}_{\gamma \gamma}  + \Omega^{11}_{\delta \delta}} \; ,
\end{align}
\end{subequations}
where the solution of Eq.~\eqref{e:tinitbccsda} must be inserted into Eq.~\eqref{e:tinitbccsdb} and where 
the operator $P_{\alpha\beta}$ permutes the two labels $\alpha$ and $\beta$. Starting from $| \Phi_{\text{HFB}} \rangle$, conditions $\Omega^{11}_{\alpha\alpha}= E_{\alpha}$ and $\Omega^{20}_{\alpha\beta}=0$ from the diagonalization of Eq.~\eqref{e:hfb} 
further simplify the initial conditions to 
\begin{subequations}
\label{e:tinitbccsdred}
\begin{align}
t_{\alpha \beta} &= 0 \; , \\
t_{\alpha \beta \gamma \delta} &= - \frac{\Omega^{40}_{\alpha \beta \gamma \delta}}
{E_{\alpha} + E_{\beta} + E_{\gamma} + E_{\delta}} \; .
\end{align}
\end{subequations}

\subsection{Particle number and other observables}

The amplitude equations (Eqs. \eqref{e:bccsd2qpex} and~\eqref{e:bccsd4qpex}) are solved iteratively while constraining the BCCSD wavefunction to carry good particle number $\text{A}$ on average. This is effected by: (i) iterating the BCCSD amplitude equations until a converged energy is obtained, (ii) computing the error in average particle number via Eq.~\eqref{deltaA}, (iii) adjusting the chemical potential to correct for the error, (iv) reinitializing quasiparticle amplitudes via Eq. \eqref{e:tinitbccsd}, and (v) returning to (i) until the targeted value of particle number is achieved at convergence. 

As discussed in Sec.~\ref{Nconstrain}, the average particle number is obtained by adding to the reference value $A^{00}$ the correction $\delta \text{A}$ computed through Eq.~\eqref{deltaA}. In the BCCSD approximation, the expectation values of the particle number and other operators are obtained from Eq. \eqref{e:genopexp}, which terminates since the truncation to singles and doubles applies also to the left reference state, i.e. $\Lambda = 
\Lambda_1 + \Lambda_2$.  Thus, one can write
\begin{eqnarray}
\label{e:opbccsd}
\Delta \text{O} &=&  \langle \Phi | \bar{O}_N | \Phi \rangle_{\text{C}}  \\
&& + \frac{1}{2} \displaystyle\sum_{\alpha \beta}  \langle \Phi | \Lambda_1 | \Phi^{\alpha \beta} \rangle \langle \Phi^{\alpha \beta} | \bar{O}_N | \Phi \rangle_{\text{C}} \nonumber \\
&& + \frac{1}{4!} \displaystyle\sum_{\alpha \beta \gamma \delta}  \langle \Phi | \Lambda_2 | \Phi^{\alpha \beta \gamma \delta} \rangle \langle \Phi^{\alpha \beta \gamma \delta} | \bar{O}_N | \Phi \rangle_{\text{C}} \nonumber \, .
\end{eqnarray}

Although it is our ambition to solve the left eigenvalue problem in the near future within the BCC framework, along with the associated equation-of-motion (EOM) method, it is not the most efficient approach for repeated evaluations.  We utilize instead an approximate implementation that consists of setting the de-excitation operator $\Lambda = {\cal T}^{\dagger}$, which is exact to first order in perturbation theory, with the potential for improvement by extending to second order in perturbation theory~\cite{shavitt09a}. This approximation is sufficient to converge the system of equations, and will be used to evaluate the variance in Sec. \ref{applications} by computing Eq. \eqref{e:vareq}.

For the operator $O=\Omega$, the terms $ \langle \Phi^{\alpha \beta} | \bar{O}_N | \Phi \rangle_{\text{C}}$ and 
$\langle \Phi^{\alpha \beta \gamma \delta} | \bar{O}_N | \Phi \rangle_{\text{C}}$ are exactly the single- and double-excitation amplitude equations.  
At convergence, we verify that the evaluation of Eq. \eqref{e:opbccsd} returns $\Delta \Omega_0$. To evaluate another operator, for instance the particle number operator $A$, one can use Eqs. \eqref{e:bccsd2qpex} and \eqref{e:bccsd4qpex}, with the normal-ordered matrix elements of $A$ as taken 
from App. \ref{1Boperator} in place of the normal-ordered grand canonical potential matrix elements. Even further, the normal-ordered matrix elements can 
be obtained from Eq. \eqref{e:me3defas} with a suitable replacement of the single particle Hamiltonian matrix elements, i.e. with $t_{pq} \rightarrow \delta_{pq},
\bar{v}_{pqrs} \rightarrow 0, \bar{w}_{pqrstu} \rightarrow 0, \lambda \rightarrow 0$ for $A$.  This procedure can be applied for 
any operator which can be similarly expressed in terms of the Hamiltonian Eq. \eqref{e:ham}, and is also used for the evaluation of $A^2$ in this work. 
  
\section{Diagrammatic method}
\label{diagrammatic}

\begin{figure}
\includegraphics[width=\columnwidth]{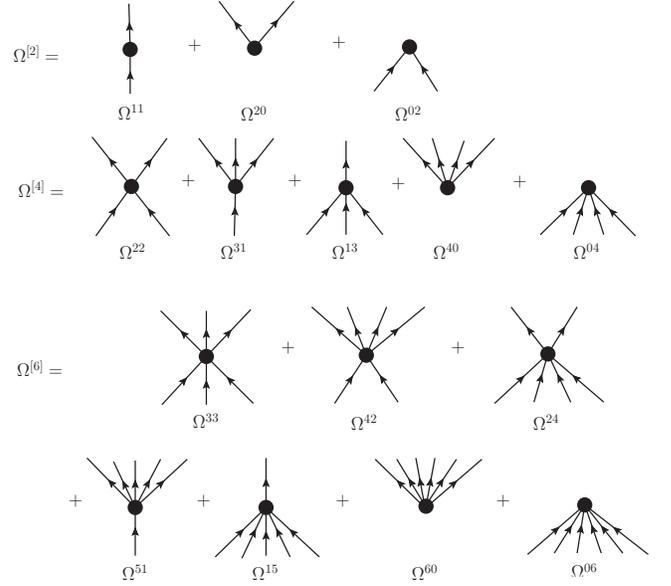}
\caption{Normal-ordered contributions to the grand canonical potential in diagrammatic form.  The first line corresponds to 
the $\Omega^{[2]}$ terms, the second line to the $\Omega^{[4]}$ terms, and the final two lines to the 
$\Omega^{[6]}$ terms, which are neglected in the present work.}
\label{f:omega}
\end{figure}

The algebraic derivation of the expanded BCC equations becomes tedious as the truncation of $\mathcal{T}$ is relaxed.  As a result, a diagrammatic technique is desired. The diagrammatic description at play in standard coupled cluster theory~\cite{shavitt09a} provides guidance for the extension to BCC theory.  In fact, the procedure is simplified with fewer diagrams at a given truncation order since particles and holes do not need to be treated separately in BCC. In agreement with the approximation used in the present work, the diagrammatic technique is constructed here by considering normal-ordered contributions $\Omega^{ij}$ up to $i+j\leq 4$. This can be eventually extended to genuine three-body terms, i.e. to $\Omega^{ij}$ terms with $i+j=6$, similarly to what was done in standard CC theory~\cite{Hagen:2007ew}.

\begin{figure}
\includegraphics[width=\columnwidth]{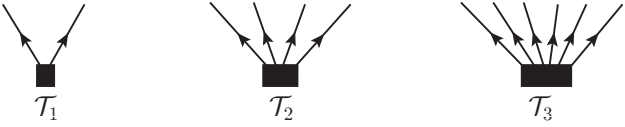}
\caption{Singly- (${\cal T}_1$),  doubly- (${\cal T}_2$) and triply- (${\cal T}_3$) excited quasiparticle cluster operators in diagrammatic form. In the present work, ${\cal T}_3$ is neglected.}
\label{f:tamps}
\end{figure}

Taking BCCSD as an example, the objective is to represent Eqs.~\eqref{e:ebccsd}, \eqref{e:bccsdt1} and \eqref{e:bccsdt2} in a diagrammatic form such that their full expanded expression given by Eqs.~\eqref{e:bccsde}, \eqref{e:bccsd2qpex} and \eqref{e:bccsd4qpex}, respectively, are obtained through the application of systematic rules while bypassing the pedestrian application of Wick's theorem. In the end, such a procedure is much more resilient against errors. To proceed, the building blocks that need to be defined are
\begin{enumerate}
\item Diagrams representing normal-ordered contributions $\Omega^{ij}$ to the grand canonical potential. The complete set of such diagrams is provided in Fig.~\ref{f:omega}. In a given diagram, one must associate the factor $\Omega^{ij}_{k_1 \ldots k_i k_{i+1} \ldots k_{i+j}}$ to the dot vertex, where $i$ denotes the number of lines representing quasiparticle creation operators (i.e. traveling out of and above the vertex) and $j$ denotes the number of lines representing quasiparticle annihilation operators (i.e. traveling into the vertex from below).  The indices $k_1 \ldots k_i$ must be assigned consecutively from the leftmost to the rightmost line above the vertex, while $k_{i+1} \ldots k_{i+j}$ must be similarly assigned consecutively for lines below the vertex.
\item Diagrams representing the n-tuple cluster amplitudes. Those diagrams are provided in Fig.~\ref{f:tamps} up to the triply-excited cluster operator ${\cal T}_3$, which is neglected in the present work. As cluster operators only contain quasiparticle creation operators, they only display lines traveling out of and above the vertex. In a given diagram, one must associate each $\mathcal{T}_m$ vertex with an amplitude $t_{k_1 \ldots k_{2m}}$, where $k_1 \ldots k_{2m}$ must be assigned consecutively from the leftmost to the rightmost line above the vertex.
\end{enumerate}
With these building blocks at hand, one needs to construct the diagrams that make up all the terms entering Eqs.~\eqref{e:ebccsd}, \eqref{e:bccsdt1} and \eqref{e:bccsdt2}. The basic rules to do so are that
\begin{enumerate}
\item All diagrams are {\it connected}, i.e. each contributing ${\cal T}_m$ operator is contracted at least once with $\Omega$.
\item Diagrams making up  Eq.~\eqref{e:ebccsd} are {\it vacuum-to-vacuum} diagrams, i.e. they are closed with no line leaving the diagram. Each diagram contributing to Eq.~\eqref{e:bccsdt1} (Eq.~\eqref{e:bccsdt2}) is {\it linked} with two (four) external lines leaving it from above.
\item For a given term in  Eqs.~\eqref{e:ebccsd}, \eqref{e:bccsdt1} and \eqref{e:bccsdt2}, construct all possible independent diagrams from the building blocks, i.e. contract the lines of $\Omega$ and of the various $\mathcal{T}_m$ in all possible ways such that the two rules above are fulfilled. Doing so typically limits which parts $\Omega^{ij}$ of $\Omega$ contribute to a given term.
\end{enumerate}
Once all the diagrams are drawn, one must compute their expressions. The rules to do so are
\begin{enumerate}
\item Label external lines with quasiparticle indices $\alpha, \beta, \ldots$ occurring in the bra of the amplitude equations.  The labeling must coincide with the left-right ordering of the indices observed in the bra. Label internal lines with different quasiparticle indices.
\item Associate the interaction vertex and the cluster amplitudes at play with the appropriate factors $\Omega^{ij}_{k_1 \ldots k_i k_{i+1} \ldots k_{i+j}}$ and $t_{k_1 \ldots k_{2m}}$, respectively.  
\item Sum over all internal line labels. 
\item Include a factor $(n!)^{-1}$ for each set of $n$ equivalent internal lines.  Equivalent internal lines are those 
which connect to identical vertices.
\item Include a factor $(\ell_m!)^{-1}$ for each set of $\ell_m$ equivalent $\mathcal{T}_{m}$ vertices.  Two $\mathcal{T}_m$ 
vertices are equivalent if they have the same number of outgoing lines $n_l \, (n_l \leq 2m)$ which terminate at the interaction vertex.
\item Provide the diagram with a sign $(-1)^{\ell_c}$, where $\ell_c$ is the number of line crossings 
in the diagram (vertices are not considered line crossings).
\item Sum over all distinct permutations $P$ of labels of inequivalent external lines, 
including a parity factor $(-1)^{\sigma(P)}$ from the signature of the permutation.  External lines are equivalent 
if and only if they connect to the same vertex.
\end{enumerate}

\begin{figure}
\includegraphics[width=\columnwidth]{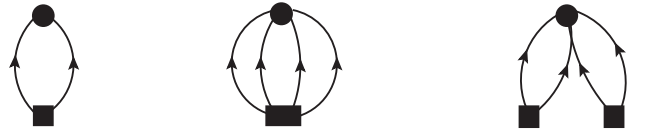}
\caption{Contributions to $\Delta\Omega_0$ in diagrammatic form. Excluding $\Omega^{[6]}$ terms, these three diagrams provide an exact form for the correction to the unperturbed energy, independent of the truncation imposed on ${\cal T}$.}
\label{f:energy}
\end{figure}

\begin{figure}
\includegraphics[width=\columnwidth]{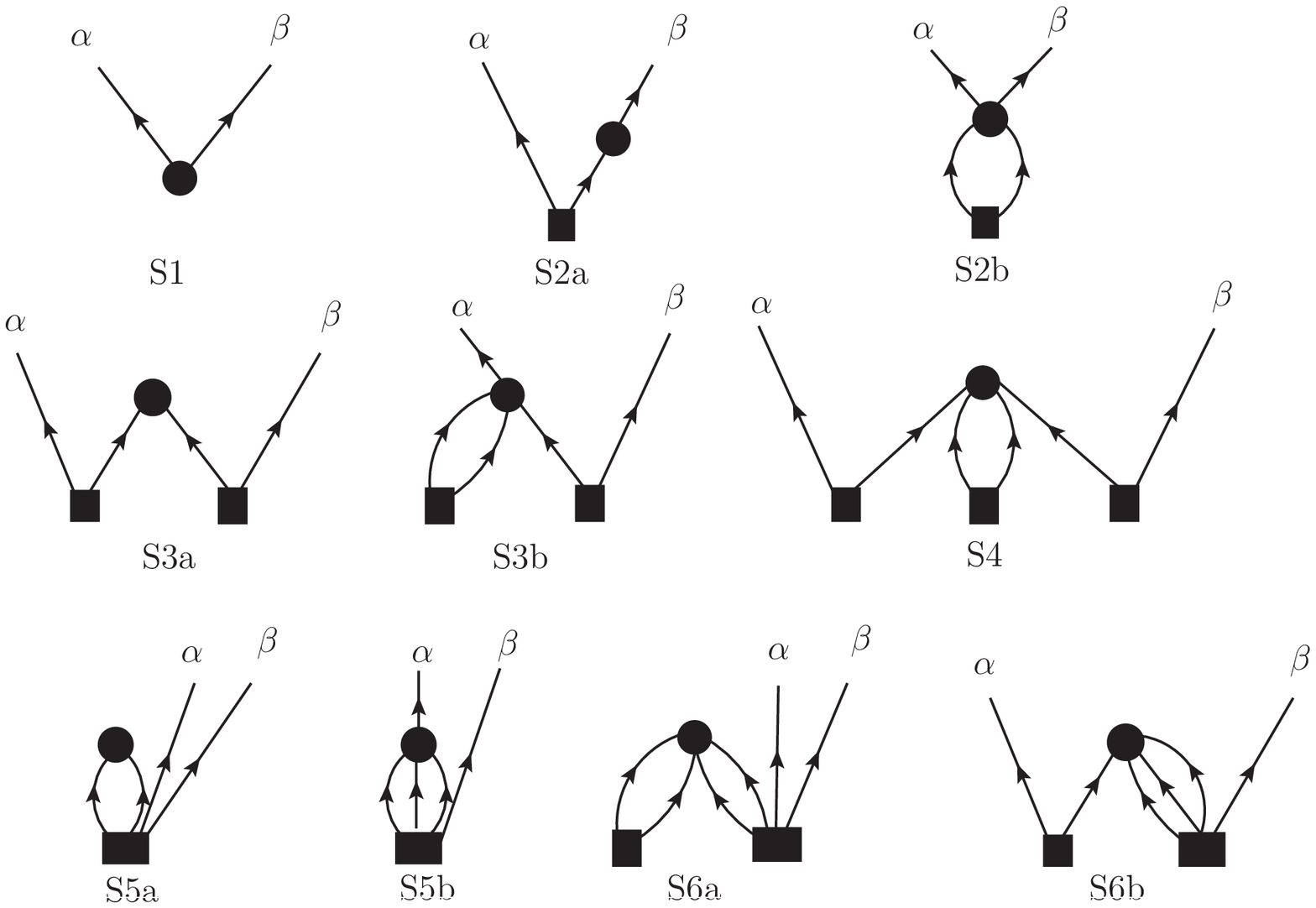}
\caption{Diagrammatic representation of the single-excitation amplitude equations in the BCCSD approximation.}
\label{f:bccsdt1}
\end{figure}

\begin{figure}
\includegraphics[width=\columnwidth]{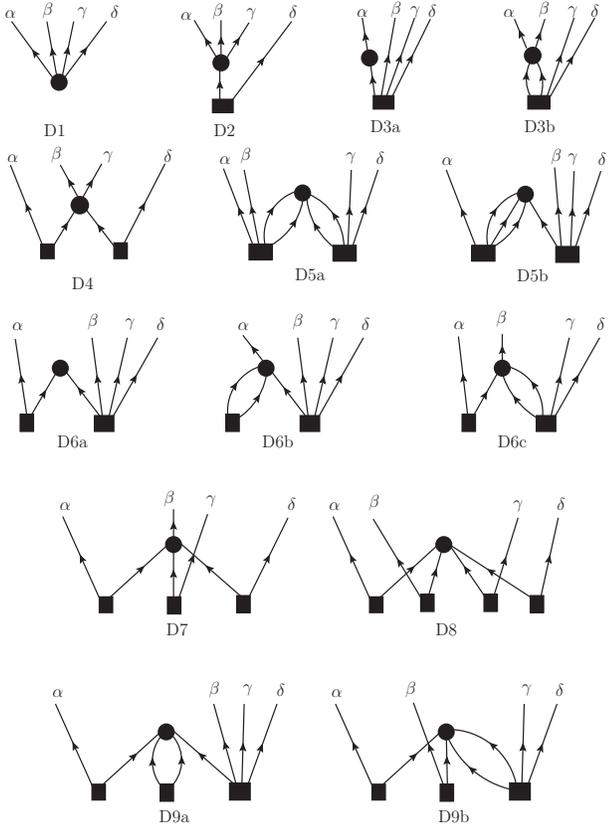}
\caption{Diagrammatic representation of the double-excitation amplitude equations in the BCCSD approximation.}
\label{f:bccsdt2}
\end{figure}

The complete set of diagrams contributing to Eqs.~\eqref{e:ebccsd}, \eqref{e:bccsdt1} and \eqref{e:bccsdt2} are given in Figs.~\ref{f:energy}, \ref{f:bccsdt1} and \ref{f:bccsdt2}, respectively. Each term in the final two diagrams is labeled, where the first character S or D refers to the single- or 
double-excitation amplitude equations, whereas the following number denotes the term in the algebraic expression to which the diagram corresponds (i.e. in Eqs. \eqref{e:bccsdt1} and \eqref{e:bccsdt2}). If there are multiple diagrams which refer to a  single algebraic term, they are labeled with a final character incremented alphabetically.  For instance, diagrams S6 of Fig. \ref{f:bccsdt1} refer to the sixth term in Eq.~\eqref{e:bccsdt1}, for which  $\Omega^{04}$ is the only term that can connect $\mathcal{T}_1$ and $\mathcal{T}_2$ to the two quasiparticle excitation level required at the top of the diagram.  However, there are two possible ways to connect them, i.e. $\mathcal{T}_1$ can either have one or two lines connected to the interaction vertex.  These possibilities are thus labeled S6a and S6b, respectively.  

Writing the algebraic result in a compact fashion requires a set of general permutation operators to handle inequivalent external lines. The permutation factor necessary for a given diagram depends on the number of external lines and their equivalence to each other. We thus employ permutation operators $P(\alpha\beta / \gamma\delta / \ldots)$ where the notation denotes that $\alpha,\beta$ and $\gamma,\delta$ are equivalent pairs, but are distinct from each other and from the remaining indices. As a result, all possible permutations among labels, except for those involving labels in the same group, are implied.  The ordering of the groups within the parentheses is irrelevant, e.g. $P(\alpha\beta/\gamma\delta/ \ldots) = P(\gamma\delta/\alpha\beta/ \ldots)$.  
In the end, the permutation operators required to express the diagrams occurring at the BCCSD level are
\begin{subequations}
\begin{eqnarray}
\label{e:perms}
P(\alpha/\beta) &\equiv& 1- P_{\alpha\beta} \\
P(\alpha\beta\gamma/\delta) &\equiv& 1 - P_{\alpha\delta} - P_{\beta\delta} - P_{\gamma\delta} \\
P(\alpha\beta/\gamma\delta) &\equiv& 1 - P_{\alpha\gamma} -P_{\alpha\delta} - P_{\beta\gamma} \nonumber \\
&& \,\,\,\, - P_{\beta\delta} + P_{\alpha\gamma}P_{\beta\delta} \\
P(\alpha/\beta\gamma/\delta) &\equiv& 1- P_{\alpha\beta} - P_{\alpha\gamma} - P_{\alpha\delta} - P_{\beta\delta} - P_{\gamma\delta} \nonumber \\
&& \,\,\,\, + P_{\alpha\beta}P_{\gamma\delta} + P_{\alpha\gamma}P_{\beta\delta}  + P_{\alpha\beta}P_{\alpha\delta} \nonumber \\
&& \,\,\,\, + P_{\alpha\gamma}P_{\alpha\delta} + P_{\beta\delta}P_{\alpha\delta} + P_{\gamma\delta}P_{\alpha\delta}.
\end{eqnarray}
\end{subequations}
In fact, from the diagrammatic rules and Diagram D8, an additional permutation operator $P(\alpha/\beta/\gamma/\delta)$ is necessary.  Based on the antisymmetry of $\Omega^{04}_{k_1 k_2 k_3 k_4}$ and the product of four quasiparticle amplitudes, the permutation operator produces 24 identical 
contributions, whose sum is the final term of Eq. \eqref{e:bccsd4qpex}.  For brevity, the form of this permutation operator has been suppressed.

\begin{figure}
\includegraphics[width=\columnwidth]{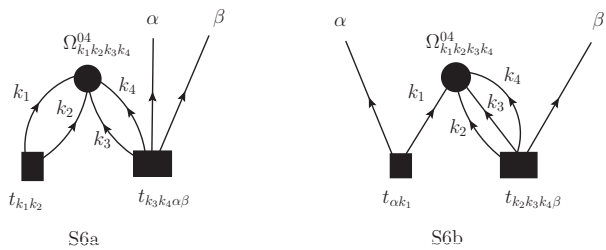}
\caption{Explicit labeling of two diagrams for the double-excitation amplitude equations in BCCSD.}
\label{f:diagex}
\end{figure}

As an illustration, we focus on diagrams S6a and S6b that are represented in complete detail in Fig.~\ref{f:diagex}, in order to provide instruction on the implementation of the rules for evaluation. Following those rules, the algebraic expression of diagram S6a is
\begin{equation}
\label{e:s6a}
\text{S}_{6\text{a}} = \frac{1}{2}\frac{1}{2} \displaystyle\sum_{k_1 k_2 k_3 k_4}
\Omega^{04}_{k_1 k_2 k_3 k_4} t_{k_1 k_2} t_{k_3 k_4 \alpha \beta} \, ,
\end{equation}
with two pairs of equivalent internal lines, no crossing lines, and two equivalent external lines. No permutation operator occurs given that the two external lines are equivalent. Similarly, Diagram S6b has the algebraic expression
\begin{equation}
\label{e:s6b}
\text{S}_{6\text{b}} = \frac{1}{3!} P(\alpha / \beta) \displaystyle\sum_{k_1 k_2 k_3 k_4}
\Omega^{04}_{k_1 k_2 k_3 k_4} t_{\alpha k_1} t_{k_2 k_3 k_4 \beta} \, ,
\end{equation}
where the factor of $(3!)^{-1}$ comes from the three equivalent internal lines. The permutation operator $P(\alpha / \beta)$ enters due to the fact that the two external lines are inequivalent. In this diagram, there are no lines crossing and therefore the sign of the diagram is positive.  These results correspond to the first three terms in the last line of Eq. \eqref{e:bccsd2qpex}, albeit in a slightly 
different ordering of indices after utilizing the antisymmetry properties of the grand canonical potential matrix elements and quasiparticle amplitudes.  In the complete description of BCCSD, there are 27 contributing diagrams as seen in  Figs.~\ref{f:energy}, \ref{f:bccsdt1} and \ref{f:bccsdt2}. The algebraic results for BCCSD obtained from Wick's theorem have been compared to those determined from the diagrammatic method to ensure the identity of the two methods. Only one method is eventually necessary such that the diagrammatic technique will be employed to set up more involved truncation schemes in the future.

\section{Proof-of-principle calculations}
\label{applications}

\subsection{Calculational scheme}

The BCC code is written in $m$-scheme starting from a spherical harmonic oscillator (HO) basis defined by its frequency $\omega$ and the number of included major shells $N_{\text{max}} \equiv$~max~$(2n+\ell)$, where $n$ is the principal quantum number and $\ell$ is the orbital angular momentum. Single-particle basis states carry quantum numbers $p \equiv (n,\pi,j,m,q)$, where $\pi=(-1)^{\ell}$ stands for the parity, $j$ for the total angular momentum, $m$ for its projection on the $z$ axis and $q$ for the projection of the isospin on the same axis. Solving the HFB problem (Eq.~\eqref{e:hfb}) within $m$-scheme provides the reference state for the BCC calculation. The normal ordering of the grand canonical potential (Eq.~\eqref{e:h3qpas}) and the BCC equations provided in Sec.~\ref{BCCSDsection} are thus implemented in the associated quasiparticle basis $\{\beta^{\dagger}_{K}\}$ carrying quantum numbers $K \equiv (k,\pi_k,m_k,q_k)$ and displaying a degeneracy according to $|m_k|$.

In both the BCCSD and BCCD approximations, the amplitude equations scale as $N^6$, where $N$ is the total number of single-particle basis states.  The scaling is slightly worse than the standard CCSD and CCD cases, 
$n_h^2 \, n_p^4$, where the basis can be split into $n_h$ hole (occupied) orbits and $n_p$ particle (unoccupied) orbits based on the underlying 
Hartree-Fock reference state.  In addition, coupled cluster codes have existed for decades, with optimized and parallelized versions available.  
While our BCC code is parallelized, significant optimization is necessary, especially in terms of the storage of quasiparticle amplitudes, which currently prevents calculations beyond $N_{\text{max}}=6$.  For example, there are 336 basis states at $N_{\text{max}}=6$, but 1820 states at $N_{\text{max}}=12$.  As the 
number of equations to be solved and matrix elements to be stored scale with the quartic power of the number of basis states, this increase is significant, requiring approximately 100TB of storage for  $N_{\text{max}}=12$.  To increase beyond $N_{\text{max}}=6$, we will either employ on-the-fly computations of matrix elements or produce an equivalent code in $J$-coupled-scheme code, where the storage required is greatly reduced. However, only the $m$-scheme code authorizes the introduction of deformation, i.e. the breaking of $SU(2)$ symmetry, to access doubly open-shell nuclei in the future. Eventually, the exact restoration of $U(1)$~\cite{duguet14a} and of $SU(2)$~\cite{Duguet:2014jja} symmetry can be handled on the basis of the same BCC $m$-scheme code.

The $m$-scheme HFB code has been benchmarked against a $J$-coupled-scheme code~\cite{soma11a} for a variety of closed- and open-shell nuclei. Being the only one of its type, the BCC code can at best be benchmarked in doubly closed-shell nuclei against an existing CC code. We have done so employing a $J$-coupled-scheme CC code~\cite{Hagen:2010gd} and checked that the results are indeed the same for $^4$He, $^{16}$O and $^{24}$O.  While it can 
be shown analytically that HFB reduces to HF in the limit of no pairing, no such analytic proof has been established to show that the application of BCC equations on top of the HF reference state will reproduce standard CC results in this limit.  In practice, however, our BCC results agree at the eV level with CC results for a variety of doubly closed-shell nuclei, model spaces, and interactions.  

The results below are based on a two-nucleon force only. The inclusion of three-nucleon forces at the normal-ordered two-body level is the goal of a forthcoming publication. We employ the chiral NNLO$_{\text{opt}}$~\cite{Ekstrom:2013kea} interaction defined with a regularization cut-off $\Lambda_{NN}= 500$~MeV and run the calculations with $N_{\text{max}}$ = 6.  Furthermore, we restrict solely to doubles excitations in the first implementation of the BCC code, with the complete demonstration of BCCSD results as obtained from the solution of Eqs.~\eqref{e:ebccsd}, \eqref{e:bccsdt1}, \eqref{e:bccsdt2} and \ref{deltaA} postponed to a future publication. 

We therefore perform BCCD calculations for the ground states of $^{16,18,20}$O, $^{18}$Ne, and $^{20}$Mg.  The doubly magic $^{16}$O nucleus provides a comparison to CCD and CCSD calculations, while the  $\text{A}=18$ nuclei will be compared to two-particle-attached equation-of-motion CCSD (2PA-EOM-CCSD) results \cite{Jansen:2011gb,piecuch2013,Jansen:2013gb}.

\begin{figure}
\includegraphics[width=\columnwidth]{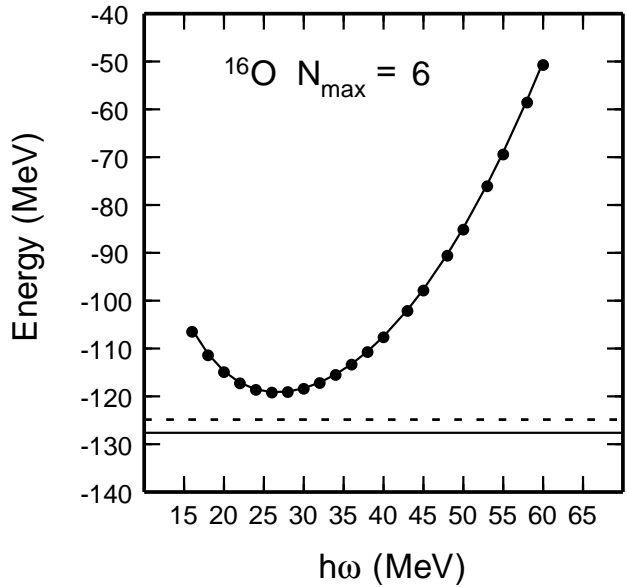}
\caption{Comparison of CCSD and CCD calculations for $^{16}$O with $N_{\text{max}}=6$ for a variety of harmonic oscillator bases given 
by $\hbar \omega$.  The points denote total energies in the CCSD approximation, with a minimum observed at -119.211 MeV for $\hbar \omega = 26 $ MeV, 
while the solid curve provides corresponding results in the more restrictive CCD approximation.  The horizontal dotted line represents the 
extrapolated energy (see text), which is indistinguishable in the CCSD and CCD approximations on the scale of the figure, while the horizontal solid line is the experimental energy.}
\label{f:o16both}
\end{figure}

In Fig. \ref{f:o16both}, CCSD calculations are compared to CCD calculations of the ground-state energy of $^{16}$O as a function of the harmonic 
oscillator basis quantum of energy (denoted by $\hbar \omega$).  In a complete model space, the energy should be constant as a function of 
$\hbar \omega$, which is far from true at $N_{\text{max}}=6$.  Regardless, the CCD results are consistently higher in energy than the CCSD results, but in reasonable agreement with a root-mean-square deviation of 315 keV for the 21 values of $\hbar \omega$ displayed in Fig. \ref{f:o16both}.  The doubles 
excitations contain the majority of the correlation energy for a two-body potential, with the optimized HF reference state of the CC equations minimizing the 
effect of singles contributions.  Therefore, the CCD approximation provides sufficient accuracy for our benchmark calculations of $^{16}$O, such that the 
truncation $\mathcal{T}=\mathcal{T}_2$ is reasonable for our proof-of-principle calculations of open-shell nuclei.  
As we likewise construct BCC equations on top of an 
optimized HFB reference state, we expect to minimize the effect of singles contributions, which can be related to a transformation of the reference state via 
the Thouless theorem~\cite{thouless61a}.

\subsection{Results}

\begin{figure}
\includegraphics[width=\columnwidth]{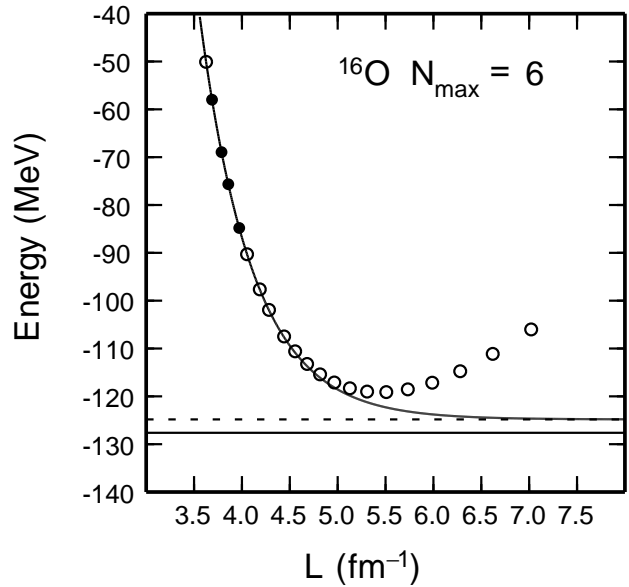}
\caption{CCD calculations for $^{16}$O with $N_{\text{max}}=6$ as a function of $L$.  The 
points denote total energies, with a minimum observed at -119.110 MeV for $\hbar \omega = 26 $ MeV.  Four points, from the calculations with 
$\hbar \omega = 50,53,55,58$ MeV, are fit to Eq. \eqref{e:extrap} and represented by the solid curve.  The horizontal dotted line is the 
extrapolated energy $E_{\infty}$, while the horizontal solid line is the experimental energy.}
\label{f:o16L2}
\end{figure}

\begin{table}
\caption{Minimum energies and associated frequencies obtained for $^{16}$O in 
different $N_{\text{max}}$ CCSD calculations.}
\begin{ruledtabular}
\begin{tabular}{lll}
$  N_{\text{max}}  $ & $\hbar \omega _{\text{min}}$ & $E_{\text{min}}$ \\
 6 & 26 & -119.211 \\
 8 & 24 & -122.776 \\
10 & 24 & -123.400 \\
12 & 22 & -123.502 \\
\end{tabular}
\label{t:ccsdo16}
\end{ruledtabular}
\end{table}

Returning to the $^{16}$O data in Fig. \ref{f:o16both}, we have verified BCCD results against CCD results in $^{16}$O, which agree with each other at the eV level.  Since the ground-state energy is not constant as a function of the basis, a minimum energy can be located as a function of the oscillator frequency.  In this example, the minimum energy -119.211 (-119.110) MeV occurs at 
$\hbar \omega = 26 $ MeV in the CCSD (CCD) approximation.  This result, however, is still underbound relative to the result 
with a complete basis.  For $^{16}$O, 
the CCSD code can be used to establish the convergence as a function of $N_{\text{max}}$.  The minimum energy for CCSD calculations up to $N_{\text{max}}=12$, along with the corresponding value of $\hbar \omega$, is shown in Tab. \ref{t:ccsdo16}.  From the convergence pattern in Tab. \ref{t:ccsdo16}, as well as the small variation observed as a function of $\hbar \omega$ in the $N_{\text{max}}=12$ calculations (not shown), we set a conservative uncertainty on the converged energy of $^{16}$O, namely $E_0=-123.5(1)$ MeV. Therefore, the missing energy due to the truncated model space at $N_{\text{max}}=6$ is approximately 3.5\% of the binding energy. Notice that, while the converged CCSD energy underbinds the experimental value of -127.619 MeV, the perturbative inclusion of triples via the $\Lambda$-CCSD(T) approximation method results in a similarly conservative converged energy of -130.3(2), thus overbound relative to experiment. 

The BCCD code currently cannot access large enough model spaces to reach energies which are converged (as a function of $N_{\text{max}}$).  In order to 
make predictions that are more accurate than 5\%, we utilize the infrared (IR) extrapolation technique \cite{furnstahl14,furnstahl14a}.  
The oscillator basis truncation effectively imposes a Dirichlet boundary condition at a radius $L$ in position space, approximated by  
\begin{equation}
L = L_{\text{eff}} \equiv \sqrt{2(N_{\text{eff}}+3/2+2)} b \, ,
\end{equation}
where $b = \sqrt{\hbar/(M \omega)}$ is the oscillator length and $N_{\text{eff}}$ is obtained by matching to the lowest eigenvalue of the squared 
momentum operator \cite{furnstahl14a}.  For $^{16}$O, the values of $N_{\text{eff}}$ are included in Tab. 1 of \textcite{furnstahl14a}.  With the 
effective radius in position space, the IR extrapolation technique, originally derived for a 
single-particle degree of freedom but now implemented also for bound many-body systems, can be implemented via the expression \cite{furnstahl14a}
\begin{equation}
\label{e:extrap}
E(L) = E_{\infty} + A_{\infty} e^{-2k_{\infty} L}.
\end{equation}
The IR extrapolation technique is reliable in nuclei around $^{16}$O, if the ultraviolet (UV) contamination is small, which can be achieved by 
using harmonic oscillator bases with UV cutoffs $\Lambda_{\text{UV}}$ greater than the momentum cutoff $\Lambda_{NN}$ of the nuclear interaction; in this case, $\Lambda_{NN} = 500$ MeV. In fact, the NNLO$_{\text{opt}}$ cutoff is not sharp, so $\Lambda_{\text{UV}}$ must be sufficiently higher than $\Lambda_{NN}$ \cite{furnstahl14a}. Erring on the side of caution, we fit Eq. \eqref{e:extrap} using energy data from $\hbar \omega = 50,53,55,58$.

\begin{figure}
\includegraphics[width=\columnwidth]{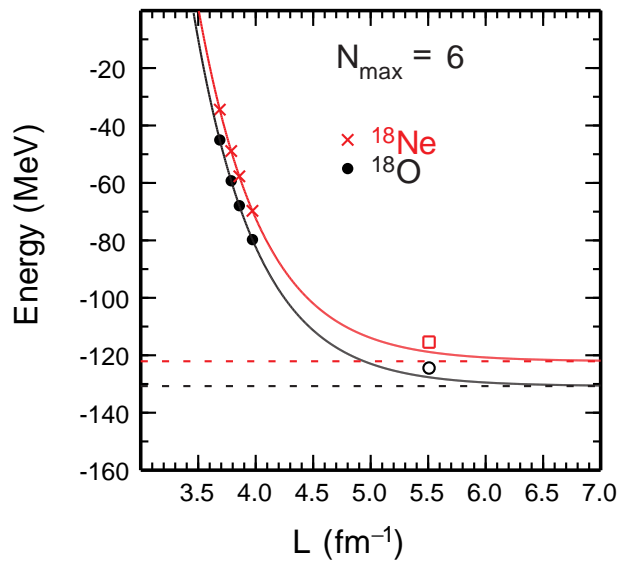}
\caption{(color online) BCCD calculations for $^{18}$O (black circles) and $^{18}$Ne (red crosses) with $N_{\text{max}}=6$ as a function of $L$.  In each nucleus, four points, from the calculations with 
$\hbar \omega = 50,53,55,58$ MeV, are fit to Eq. \eqref{e:extrap} and represented by the solid curve.  The open symbols correspond to results at $\hbar \omega = 26$ MeV.  The horizontal dotted line is the 
extrapolated energy $E_{\infty}$. See text for additional details.}
\label{f:irconv18}
\end{figure}

\begin{figure}
\includegraphics[width=\columnwidth]{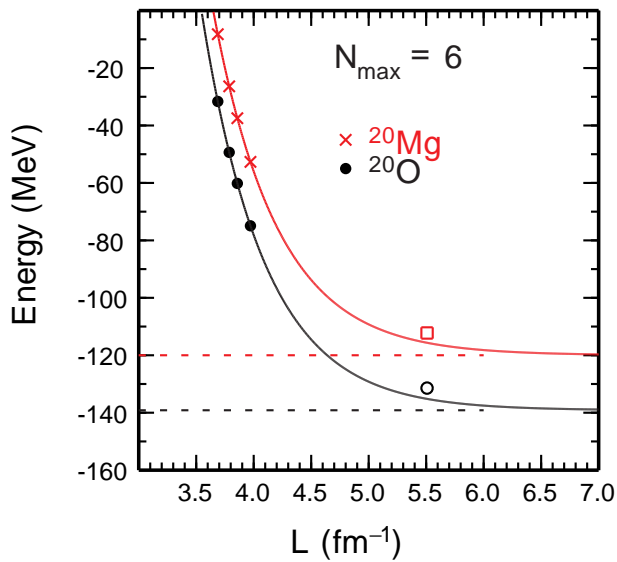}
\caption{(color online) BCCD calculations for $^{20}$O (black circles) and $^{20}$Mg (red crosses).  See caption to Fig. \ref{f:irconv18} for details.}
\label{f:irconv20}
\end{figure}

In Fig. \ref{f:o16L2}, we plot the same CCD data shown in Fig. \ref{f:o16both}, but now as a function of $L$ instead of $\hbar \omega$.  The filled circles correspond to points used in the fit for the IR extrapolation, given by Eq. \eqref{e:extrap}, while the solid curve displays the function determined by a least-squares fitting routine.  The fit is perfect at the keV level, and results in a parameter $E_{\infty}$ which corresponds to the energy in the infinite basis.  We thus obtain the extrapolated value for the energy denoted by the horizontal dashed line, $E_{\infty} = -124.821$ MeV.  In comparison to the converged energy of $^{16}$O, -123.5(1) MeV, we observe 1.3 MeV overbinding from the extrapolation procedure.  This is consistent with the results of \cite{furnstahl14a} for $N_{\text{max}}=6$ \cite{kylepc}, whereas extrapolations from $N_{\text{max}}=8,10,12$ do not lead to overbinding.  Possible explanations for this imperfect extrapolation are insufficient decoupling of the center of mass due to the small model space, or a peculiarity of the CCSD calculation at $N_{\text{max}}=6$ \cite{furnstahl14a}.  With the knowledge gleaned from CC calculations of $^{16}$O, we produce BCCD calculations of $^{16,18,20}$O, $^{18}$Ne, $^{20}$Mg for $\hbar \omega =26,50,53,55,58$.  The lowest frequency is used as an approximate value to obtain the minimum energy for the $N_{\text{max}}=6$ calculations without performing time-consuming calculations to determine the minimum for each nucleus.  The high frequencies provide the data to fit the IR extrapolation formula (Eq. \eqref{e:extrap}) to find an extrapolated ground-state energy for the NNLO$_{\text{opt}}$ interaction.  The BCCD results for $^{16}$O reproduce the CCD results at the eV level, including the extrapolated result.  The $\text{A}=18$ and $\text{A}=20$ results are displayed in Figs.~\ref{f:irconv18} and~\ref{f:irconv20}, respectively, while numerical values of interest for all five nuclei can be found in Tab. \ref{t:comp}.

While CC calculations require a doubly closed-shell nucleus, there are extensions via the equation-of-motion method to compute nuclei with two nucleons
added or removed.  Therefore, BCCD results for $\text{A}=18$ nuclei can be compared to those obtained from extensions of the standard CCSD method, as shown in Tab. \ref{t:comp}. For these calculations, we used the two-particles-attached equation-of-motion (2PA-EOM) method \cite{Jansen:2011gb,piecuch2013,Jansen:2013gb} including up to $3p\!-\!1h$ excitations on top of $^{16}$O.  While the $\text{A}=20$ nuclei could in principle be accessed by two-particle-removed equation-of-motion CCSD \cite{Jansen:2011gb} relative to $^{22}$O and $^{22}$Si, since these closed-subshell nuclei are accessible via CCSD calculations, one finds significant underbinding when including up to $1p\!-\!3h$ excitations.   For $N_{\text{max}}=6$ calculations, the 2PA-EOM-CCSD results are lower in energy than the 
BCCD results by 2.04 MeV for $^{18}$O and 2.51 MeV for $^{18}$Ne. The effect of singles contributions, which have been excluded from the BCC calculations, is expected to remain on the order of 100 keV such that 2PA-EOM-CCSD is genuinely lower in energy.  This is not surprising given that nuclei in the very vicinity of a closed shell are those for which the benefit provided by the breaking of $U(1)$ symmetry is actually overtaken by the associated shortcoming of not having an exact eigenstate of the particle-number operator~\cite{Henderson:2014vka}, i.e. this constitutes a regime for which the exact restoration of symmetry~\cite{duguet14a} is critical. The present comparison of BCCD (without symmetry restoration) and 2PA-EOM-CCSD results is meant to provide a reference corresponding to the worst case scenario.\footnote{The improved performance of 2PA-EOM-CCSD calculations compared to symmetry-unrestricted single-reference CCSD calculations (without symmetry restoration) was similarly seen for $SU(2)$ symmetry in $^{6}$He, where 2PA-EOM-CCSD provided significant more binding~\cite{Jansen:2011gb}.}

\begin{table}
\caption{Compiled results for proof-of-principle BCCD calculations with $N_{\text{max}}=6$, including both the approximate minimum energy
(taken from a calculation with $\hbar \omega = 26 $ MeV) and the extrapolated energy for an infinite model space ($E_{\infty}$) via Eq. \ref{e:extrap}. 
Also included for comparison are CCSD calculations for $^{16}$O and 2PA-EOM-CCSD results for $^{18}$O and $^{18}$Ne with $\hbar \omega = 
26 $ MeV, as well as 
experimental values \cite{ame2012}.  Note that the BCCD calculations for $^{16}$O, both $E_{\text{min}}$ and $E_{\infty}$, reproduce exactly the 
corresponding CCD calculations. }
\begin{ruledtabular}
\begin{tabular}{llllll}
Nucleus & $E_{\text{min}}$ & $E^{\text{CCSD}}_{N_{\text{max}}=6} $ & $E_{\infty}$ &  $E^{\text{CCSD}}_{N_{\text{max}}=12}$ & $E^{\text{exp}}$  \\
 $^{16}$O & -119.110 & -119.211 & -124.821 & -123.453 & -127.619 \\
 $^{18}$O & -124.440 & -126.476 & -130.738 & -132.990 & -139.808 \\
 $^{20}$O & -131.428 & n/a & -139.144 & n/a &  -151.371 \\
 $^{18}$Ne & -115.413 & -117.927 & -122.089 & -124.850 & -132.143 \\
 $^{20}$Mg & -112.237 & n/a & -119.996 & n/a & -134.480 \\
\end{tabular}
\label{t:comp}
\end{ruledtabular}
\end{table}

\begin{table}
\caption{Variance in particle number from the solution to HFB equations and BCCD equations,
taken from a calculation with $N_{\text{max}}=6$ and $\hbar \omega = 26 $ MeV.}
\begin{ruledtabular}
\begin{tabular}{lll}
Nucleus &  $\Delta \text{A}^2_{\text{HFB}}$ & $\Delta \text{A}^2_{\text{BCCD}}$\\
 $^{16}$O & 0.000 & 0.000 \\
 $^{18}$O & 2.775 & 2.814 \\
 $^{20}$O & 2.888 & 3.398 \\ 
 $^{18}$Ne & 2.765 & 2.761 \\
 $^{20}$Mg & 2.859 & 2.547 \\
\end{tabular}
\label{t:var}
\end{ruledtabular}
\end{table}

While NNLO$_{\text{opt}}$ reasonably reproduces the binding energies of oxygen isotopes \cite{Ekstrom:2013kea}, and might therefore be 
expected to reproduce all five nuclei presented here, the primary motivation of this section is not to compare our results to experiment.  
Future developments of the code are needed in order to go beyond calculations at $N_{\text{max}}=6$, not only to improve the extrapolation \cite{furnstahl14a} but also to ensure that this extrapolation holds for the quasiparticle basis.  In addition, the truncation of the 
quasiparticle excitation operator at the doubles level is too restrictive.  
For instance, including triples non-iteratively in the standard coupled cluster framework via the 
$\Lambda$-CCSD(T) method~\cite{taube08a} lowers the total energy by more than six MeV for the nuclei considered here, 
better reproducing experiment in all cases.  The triples correction must therefore be included at least in a non-iterative way.
Additionally, even though the optimized two-body force utilized here reasonably reproduces ground-state properties of nuclei in the vicinity of $^{16}$O, 
the machinery to include three-body forces, at least at the normal-ordered two-body level, must be developed in order to make 
reliable theoretical predictions throughout the nuclear chart.  Future publications will 
address progress along these fronts.

Finally, as discussed in Section \ref{Nconstrain}, one should monitor the breaking of particle-number symmetry by computing the variance associated with 
the operator $A$ using  Eq. \eqref{e:vareq}.  The results are shown in Tab. \ref{t:var}.  For the five nuclei calculated here, the variance in particle number obtained at the HFB level is nearly constant.  The inclusion of additional correlations can either increase or decrease the variance, based on the nucleus of interest, but remains reasonably similar to the HFB variance, providing confidence in the applicability of the symmetry-breaking BCC equations.  Nevertheless, the behavior of the variance must be studied further, especially with respect to an increase in the model space size and based on the inclusion of singles and triples excitations. Eventually, the spontaneously-broken symmetry must be restored for a proper comparison to physical (i.e. symmetry-conserving) nuclei, for which the implementation discussed by \textcite{duguet14a} will be applied.

\section{Conclusions}
\label{conclusions}

The Bogoliubov coupled cluster theory has been formulated as a way to extend single-reference coupled cluster techniques to the description of genuinely open-shell nuclei. The rationale behind this extension is the representation of the exact ground-state wavefunction of even-even nuclei as the exponential of a quasiparticle excitation cluster operator acting on a Bogoliubov reference state. As such, BCC theory exploits the spontaneous breaking of $U(1)$ symmetry associated with particle-number conservation to overcome the degenerate character of open-shell systems. Thus, the potential span of ab initio coupled cluster calculations based on single-reference techniques is increased tremendously. 

Equations for the ground-state energy and the cluster amplitudes have been derived at the singles and doubles level (BCCSD) both algebraically and diagrammatically. The equations have been implemented in the BCCD approximation in an $m$-scheme code based on a harmonic oscillator basis, with results for a set of light doubly closed-shell nuclei validated against CCD results. The numerical scaling of the method is polynomial and goes as $N^6$ in both the BCCD and BCCSD approximations, where $N$ is the total number of single-particle basis states.

The results of the first proof-of-principle calculations have been reported for five even-even $sd$-shell nuclei in the BCCD approximation. The breaking of $U(1)$ symmetry has been monitored by computing the variance associated with the particle-number operator. The newly developed many-body formalism offers a wealth of potential applications and further extensions dedicated to the ab initio description of ground- and excited-states of open-shell nuclei. Short term extensions include the implementation of three-nucleon forces at the normal-ordered two-body level. Mid-term extensions include the development of approximate triple corrections and of the equation-of-motion methodology to treat both excited states and odd nuclei.  One can also envision calculations of doubly open-shell nuclei via the further breaking of $SU(2)$ symmetry associated with angular momentum conservation. Longer-term extensions include the exact restoration of $U(1)$~\cite{duguet14a} and $SU(2)$~\cite{Duguet:2014jja} symmetries.

\section*{Acknowledgements}

The authors would like to thank V. Som\`a for his aid in benchmarking and troubleshooting the m-scheme Hartree-Fock-Bogoliubov code, T. Papenbrock and K.A. Wendt for discussions and data relevant to extrapolations and coupled cluster methods, and T. Henderson for useful discussions regarding particle-number variance and convergence in coupled cluster methods with pairing.  A. S. acknowledges  support from Espace de Structure Nucl\'eaire Th\'eorique (ESNT).  This work was supported in part by the U.S. Department of Energy (Oak Ridge National Laboratory), under Grant Nos. DEFG02-96ER40963 (University of Tennessee), DE-SC0008499 (NUCLEI Sci-DAC collaboration), and the Field Work Proposal ERKBP57. Computer time was provided by the Innovative and Novel Computational Impact on Theory and Experiment (INCITE) program.
This research used resources of the Oak Ridge Leadership Computing Facility located at Oak Ridge National Laboratory, which is supported by the Office of Science of the Department of Energy under Contract No. DE-AC05-00OR22725.  

\begin{appendix}

\section{Normal-ordered matrix elements}

\subsection{Grand canonical potential}
\label{s:hamas}

As the $\Omega^{[6]}$ terms are not considered for practical applications at this point, the matrix elements $\Omega^{ij}_{k_1 k_2 k_3 k_4 k_5 k_6}$, with $i+j=6$, are excluded for brevity. The grand canonical potential of Eq. \eqref{e:h3qpas}, up to and including $\Omega^{[4]}$, displays fully antisymmetrized matrix elements whose explicit expressions in terms of matrix elements of the kinetic energy plus two- and three-body interactions, as well as of $U$ and $V$ matrices defining the reference Bogoliubov state, are
\begin{widetext}
\begin{subequations}
\label{e:me3defas}
\begin{align}
\label{e:me3defasa}
\Omega^{00} & = \text{Tr}\Big[T \rho + \frac{1}{2} \Gamma^{2N} \rho + \frac{1}{3}
\Gamma^{3N} \rho - \frac{1}{2}  \Delta^{2N} \kappa^{*} + \frac{1}{3} \Delta^{3N} \kappa^{*} -\lambda \rho\Big]\\
\label{e:me3defasb}
 \Omega^{11}_{k_1 k_2}  & =  [U^{\dagger} h U - V^{\dagger} h^{T} V + 
 U^{\dagger} \Delta V - V^{\dagger} \Delta ^{*} U ]_{k_1 k_2} \\
\label{e:me3defasc}
 \Omega^{20}_{k_1 k_2} & = [U^{\dagger} h V^{*} - V^{\dagger} h^{T}U^{*} + 
U^{\dagger} \Delta U^{*} - V^{\dagger} \Delta^{*} V^{*}]_{k_1 k_2} \\
\label{e:me3defasd}
 \Omega^{02}_{k_1 k_2} & = [-V^{T} h U + U^{T} h^{T} V - 
 V^{T} \Delta V + U^{T} \Delta ^{*} U]_{k_1 k_2} \\
\label{e:me3defase}
 \nonumber 
  \Omega^{22}_{k_1 k_2 k_3 k_4} &= \displaystyle\sum_{l_1 l_2 l_3 l_4}\Big[ \Theta_{l_1 l_2 l_3 l_4} \Big(
 U^{*}_{l_1 k_1}U^{*}_{l_2 k_2}U_{l_3 k_3}U_{l_4 k_4} + V^{*}_{l_3 k_1}  V^{*}_{l_4 k_2}V_{l_1 k_3}V_{l_2 k_4}  
 + U^{*}_{l_1 k_1}V^{*}_{l_4 k_2}V_{l_2 k_3}U_{l_3 k_4} \\
  \nonumber & \hspace{0.5cm} - V^{*}_{l_4 k_1}U^{*}_{l_1 k_2}V_{l_2 k_3}U_{l_3 k_4} 
- U^{*}_{l_1 k_1}V^{*}_{l_4 k_2}U_{l_3 k_3}V_{l_2 k_4} + V^{*}_{l_4 k_1}U^{*}_{l_1 k_2}U_{l_3 k_3}V_{l_2 k_4}\Big)\\
  \nonumber & \hspace{0.5cm} + \Xi_{l_1 l_2 l_3 l_4} 
  (U^{*}_{l_1 k_1}U^{*}_{l_2 k_2}U_{l_4 k_3}V_{l_3 k_4} + U^{*}_{l_1 k_1}V^{*}_{l_4 k_2}V_{l_3 k_3}V_{l_2 k_4}
  -U^{*}_{l_1 k_1}U^{*}_{l_2 k_2}V_{l_3 k_3}U_{l_4 k_4} - V^{*}_{l_4 k_1}U^{*}_{l_1 k_2}V_{l_3 k_3}V_{l_2 k_4}) \\
  & \hspace{0.5cm} - \Xi^{*}_{l_1 l_2 l_3 l_4}(V^{*}_{l_3 k_1}U^{*}_{l_4 k_2}U_{l_1 k_3}U_{l_2 k_4} +
  V^{*}_{l_3 k_1}V^{*}_{l_2 k_2}V_{l_4 k_3}U_{l_1 k_4} - U^{*}_{l_4 k_1}V^{*}_{l_3 k_2}U_{l_1 k_3}U_{l_2 k_4} -
  V^{*}_{l_3 k_1}V^{*}_{l_2 k_2}V_{l_4 k_4}U_{l_1 k_3}) \Big] \\
\label{e:me3defasf}
    \nonumber
 \Omega^{31}_{k_1 k_2 k_3 k_4} &=  \displaystyle\sum_{l_1 l_2 l_3 l_4}\Big[\Theta_{l_1 l_2 l_3 l_4}
\Big(U^{*}_{l_1 k_1}V^{*}_{l_4 k_2}V^{*}_{l_3 k_3}V_{l_2 k_4} - V^{*}_{l_4 k_1}U^{*}_{l_1 k_2}
V^{*}_{l_3 k_3}V_{l_2 k_4} - V^{*}_{l_3 k_1}V^{*}_{l_4 k_2}U^{*}_{l_1 k_3}V_{l_2 k_4} \\
\nonumber & \hspace{0.5cm} + V^{*}_{l_3 k_1}U^{*}_{l_2 k_2}U^{*}_{l_1 k_3}U_{l_4 k_4} 
-U^{*}_{l_2 k_1}V^{*}_{l_3 k_2}U^{*}_{l_1 k_3}U_{l_4 k_4} - 
U^{*}_{l_1 k_1}U^{*}_{l_2 k_2}V^{*}_{l_3 k_3}U_{l_4 k_4}\Big)   \\
  \nonumber & \hspace{0.5cm} + \Xi_{l_1 l_2 l_3 l_4} \Big(U^{*}_{l_1 k_1}U^{*}_{l_2 k_2}U^{*}_{l_3 k_3}U_{l_4 k_4} 
  + V^{*}_{l_4 k_1}U^{*}_{l_2 k_2}U^{*}_{l_1 k_3}V_{l_3 k_4} - U^{*}_{l_2 k_1}V^{*}_{l_4 k_2}U^{*}_{l_1 k_3}V_{l_3 k_4} + U^{*}_{l_2 k_1}U^{*}_{l_1 k_2}V^{*}_{l_4 k_3}V_{l_3 k_4}\Big)\\
 & \hspace{0.5cm} + \Xi^{*}_{l_1 l_2 l_3 l_4}\Big(U^{*}_{l_4 k_1}V^{*}_{l_3 k_2}V^{*}_{l_2 k_3}U_{l_1 k_4} -
 V^{*}_{l_3 k_1}U^{*}_{l_4 k_2}V^{*}_{l_2 k_3}U_{l_1 k_4} + V^{*}_{l_3 k_1}V^{*}_{l_2 k_2}U^{*}_{l_4 k_3}U_{l_1 k_4}
 - V^{*}_{l_3 k_1}V^{*}_{l_2 k_2}V^{*}_{l_1 k_3}V_{l_4 k_4}\Big) \Big]\\
\label{e:me3defasg}
  \nonumber
  \Omega^{13}_{k_1 k_2 k_3 k_4} &=  \displaystyle\sum_{l_1 l_2 l_3 l_4}\Big[\Theta_{l_1 l_2 l_3 l_4} \Big(
V^{*}_{l_4 k_1}U_{l_3 k_2}V_{l_2 k_3}V_{l_1 k_4} - V^{*}_{l_4 k_1}V_{l_2 k_2}U_{l_3 k_3}V_{l_1 k_4} 
- V^{*}_{l_4 k_1}V_{l_1 k_2}V_{l_2 k_3}U_{l_3 k_4} \\
\nonumber & \hspace{0.5cm} + U^{*}_{l_1 k_1}V_{l_2 k_2}U_{l_3 k_3}U_{l_4 k_4} - 
U^{*}_{l_1 k_1}U_{l_3 k_2}V_{l_2 k_3}U_{l_4 k_4} + U^{*}_{l_1 k_1}U_{l_3 k_2}U_{l_4 k_3}V_{l_2 k_4}\Big)   \\
  \nonumber & \hspace{0.5cm} + \Xi_{l_1 l_2 l_3 l_4} \Big(U^{*}_{l_1 k_1}V_{l_2 k_2}V_{l_3 k_3}U_{l_4 k_4} - 
  V^{*}_{l_4 k_1}V_{l_1 k_2}V_{l_2 k_3}V_{l_3 k_4} + U^{*}_{l_1 k_1}U_{l_4 k_2}V_{l_2 k_3}V_{l_3 k_4} 
  - U^{*}_{l_1 k_1}V_{l_2 k_2}U_{l_4 k_3}V_{l_3 k_4}\Big) \\
 & \hspace{0.5cm} + \Xi^{*}_{l_1 l_2 l_3 l_4}\Big(V^{*}_{l_3 k_1}V_{l_4 k_2}U_{l_1 k_3}U_{l_2 k_4} -
V^{*}_{l_3 k_1}U_{l_1 k_2}V_{l_4 k_3}U_{l_2 k_4}  + V^{*}_{l_3 k_1}U_{l_1 k_2}U_{l_2 k_3}V_{l_4 k_4}
 - U^{*}_{l_4 k_1}U_{l_1 k_2}U_{l_2 k_3}U_{l_3 k_4}\Big) \Big]\\ 
 \label{e:me3defash}
  \nonumber
   \Omega^{40}_{k_1 k_2 k_3 k_4} &=  \displaystyle\sum_{l_1 l_2 l_3 l_4}\Big[ \Theta_{l_1 l_2 l_3 l_4} 
 \Big(U^{*}_{l_1 k_1}U^{*}_{l_2 k_2}V^{*}_{l_4 k_3}V^{*}_{l_3 k_4} - U^{*}_{l_1 k_1}V^{*}_{l_4 k_2}
 U^{*}_{l_2 k_3}V^{*}_{l_3 k_4} - V^{*}_{l_4 k_1}U^{*}_{l_2 k_2}U^{*}_{l_1 k_3}V^{*}_{l_3 k_4} \\
 \nonumber & \hspace{0.5cm}+ U^{*}_{l_1 k_1}V^{*}_{l_4 k_2}V^{*}_{l_3 k_3}U^{*}_{l_2 k_4} + V^{*}_{l_4 k_1}
 U^{*}_{l_2 k_2}V^{*}_{l_3 k_3}U^{*}_{l_1 k_4} + V^{*}_{l_4 k_1}V^{*}_{l_3 k_2}U^{*}_{l_1 k_3}U^{*}_{l_2 k_4} \Big) \\
 \nonumber & \hspace{0.5cm} + \Xi_{l_1 l_2 l_3 l_4} \Big(U^{*}_{l_1 k_1}U^{*}_{l_2 k_2}U^{*}_{l_3 k_3}
 V^{*}_{l_4 k_4} - U^{*}_{l_1 k_1}U^{*}_{l_2 k_2}V^{*}_{l_4 k_3}U^{*}_{l_3 k_4} + U^{*}_{l_1 k_1}V^{*}_{l_4 k_2}
 U^{*}_{l_2 k_3}U^{*}_{l_3 k_4} - V^{*}_{l_4 k_1}U^{*}_{l_1 k_2}U^{*}_{l_2 k_3}U^{*}_{l_3 k_4} \Big) \\
  &\hspace{0.5cm} + \Xi^{*}_{l_1 l_2 l_3 l_4} \Big(V^{*}_{l_1 k_1}V^{*}_{l_2 k_2}V^{*}_{l_3 k_3}
 U^{*}_{l_4 k_4} -  V^{*}_{l_1 k_1}V^{*}_{l_2 k_2}U^{*}_{l_4 k_3}V^{*}_{l_3 k_4} 
 +V^{*}_{l_1 k_1}U^{*}_{l_4 k_2}V^{*}_{l_2 k_3}V^{*}_{l_3 k_4} - 
 U^{*}_{l_4 k_1}V^{*}_{l_1 k_2}V^{*}_{l_2 k_3}U^{*}_{l_3 k_4}\Big) \Big] \\
\label{e:me3defasi}
  \nonumber
   \Omega^{04}_{k_1 k_2 k_3 k_4} &=  \displaystyle\sum_{l_1 l_2 l_3 l_4}\Big[ \Theta_{l_1 l_2 l_3 l_4} \Big
 (U_{l_3 k_1}U_{l_4 k_2}V_{l_2 k_3}V_{l_1 k_4} - U_{l_3 k_1}V_{l_2 k_2}U_{l_4 k_3}V_{l_1 k_4} 
+ U_{l_3 k_1}V_{l_2 k_2}V_{l_1 k_3}U_{l_4 k_4} \\
 \nonumber & \hspace{0.5cm}- V_{l_2 k_1}U_{l_3 k_2}V_{l_1 k_3}U_{l_4 k_4} 
+ V_{l_2 k_1}V_{l_1 k_2}U_{l_3 k_3}U_{l_4 k_4} + V_{l_2 k_1}
 U_{l_3 k_2}U_{l_4 k_3}V_{l_1 k_4} \Big ) \\
 \nonumber & \hspace{0.5cm} + \Xi_{l_1 l_2 l_3 l_4} \Big(
 V_{l_1 k_1}V_{l_2 k_2}V_{l_3 k_3}U_{l_4 k_4} - V_{l_1 k_1}V_{l_2 k_2}U_{l_4 k_3}
 V_{l_3 k_4} + V_{l_1 k_1}U_{l_4 k_2}V_{l_2 k_3}V_{l_3 k_4} - U_{l_4 k_1}V_{l_1 k_2}V_{l_2 k_3}V_{l_3 k_4}\Big) \\
 & \hspace{0.5cm} + \Xi^{*}_{l_1 l_2 l_3 l_4} \Big(V_{l_4 k_1}U_{l_3 k_2}U_{l_2 k_3}U_{l_1 k_4} - 
 U_{l_3 k_1}V_{l_4 k_2}U_{l_2 k_3}U_{l_1 k_4} +U_{l_3 k_1}U_{l_2 k_2}V_{l_4 k_3}U_{l_1 k_4} - 
 U_{l_3 k_1}U_{l_2 k_2}U_{l_1 k_3}V_{l_4 k_4}\Big) \Big]. 
\end{align}
\end{subequations}
The above expressions make use of four one- and two-body operators whose matrix elements are given in an arbitrary single-particle basis by
\begin{subequations}
\label{variousdefinitions}
\begin{eqnarray}
h_{pq} &\equiv& t_{pq}- \lambda \, \delta_{pq} + \Gamma^{2N}_{pq} + \Gamma^{3N}_{pq}   \\
&=& t_{pq}- \lambda \, \delta_{pq} + \sum_{rs} \bar{v}_{psqr} \rho_{rs}  + \frac{1}{2} \sum_{rstu} \bar{w}_{prsqtu} \Big( \rho_{us} \rho_{tr} + 
\frac{1}{2} \kappa^{*}_{rs}\kappa_{tu} \Big)   \, , \\
\Delta_{pq} &\equiv& \Delta^{2N}_{pq} + \Delta^{3N}_{pq} \\
&=& \frac{1}{2} \sum _{rs} \bar{v}_{pqrs} \kappa _{rs} + \frac{1}{2} \sum_{rstu} \bar{w}_{rpqstu} \rho_{sr} \kappa_{tu} \, , \\
\Theta_{pqrs} &\equiv& \bar{v}_{pqrs} + \sum_{tu} \bar{w}_{pqtrsu} \rho_{ut} \, , \\
\Xi_{pqrs} &\equiv& \frac{1}{2} \sum_{tu} \bar{w}_{pqrstu} \kappa_{tu} \, .
\end{eqnarray}
\end{subequations}
\end{widetext}
It is easy to verify the following properties
\begin{subequations}
\begin{align}
\Gamma^{2N}_{pq} &= \Gamma^{2N*}_{qp} \, , \\
\Gamma^{3N}_{pq} &= \Gamma^{3N*}_{qp} \, , \\
\Delta^{2N}_{pq} &= - \Delta^{2N}_{qp} \, , \\
\Delta^{3N}_{pq} &= - \Delta^{3N}_{qp} \, , \\
\Theta_{pqrs} &= - \Theta_{pqsr} = \Theta_{qpsr} = - \Theta_{qprs} \, , \\
\Theta_{pqrs} &= \Theta^{*}_{rspq}, \\
\Xi_{pqrs} &= -\Xi_{qprs} = \Xi_{qrps} = - \Xi_{prqs} = \Xi_{rpqs} = - \Xi_{rqps} \, .
\end{align}
\end{subequations}
From these relations, it is straightforward to show that the matrix elements of the normal-ordered grand canonical potential exhibit the following behavior under hermitian conjugation
\begin{subequations}
\label{e:me3sym}
\begin{align}
\label{e:me3syma}
\Omega^{11}_{k_1 k_2} &= \Omega^{11*}_{k_2 k_1} \, ,\\
\label{e:me3symc}
\Omega^{20}_{k_1 k_2} &= \Omega^{02*}_{k_1 k_2} \, , \\
\label{e:me3symb}
\Omega^{22}_{k_1 k_2 k_3 k_4} &= \Omega^{22*}_{k_3 k_4 k_1 k_2} \, , \\
\label{e:me3symf}
\Omega^{31}_{k_1 k_2 k_3 k_4} &= \Omega^{13*}_{k_4 k_1 k_2 k_3} \, , \\
\label{e:me3symj}
\Omega^{40}_{k_1 k_2 k_3 k_4} &= \Omega^{04*}_{k_1 k_2 k_3 k_4} \, .
\end{align}
\end{subequations}

\subsection{Generic one-body operator}
\label{1Boperator}

We define a generic one-body operator
\begin{eqnarray}
O &\equiv& \sum_{pq} o_{pq} c^{\dagger}_p c_q \, .
\end{eqnarray}
Its normal ordered form with respect to $| \Phi \rangle$ is given by
\begin{subequations}
\begin{eqnarray}
O &\equiv& O^{[0]} + O^{[2]}  \\
&=& O^{00}  \\
&& + \frac{1}{1!} \sum_{k_1 k_2} O^{11}_{k_1 k_2}\beta^{\dagger}_{k_1} \beta_{k_2} \\
&& + \frac{1}{2!}\displaystyle\sum_{k_1 k_2} \Big \{O^{20}_{k_1 k_2} \beta^{\dagger}_{k_1}
 \beta^{\dagger}_{k_2} + O^{02}_{k_1 k_2}   \beta_{k_2} \beta_{k_1} \Big \} \, ,
\end{eqnarray}
\end{subequations}
with the matrix elements given by
\begin{subequations}
\begin{align}
O^{00} &=  \text{Tr}\big[ o \rho\big] \, ,\\
O^{11}_{k_1 k_2} &=  [U^{\dagger}oU - V^{\dagger}o^TV]_{k_1 k_2} \, ,\\
O^{20}_{k_1 k_2} &=  [U^{\dagger}oV^{*} - V^{\dagger}o^TU^{*}]_{k_1 k_2} \, , \\
O^{02}_{k_1 k_2} &=  [-V^{T}oU + U^{T}o^TV]_{k_1 k_2}  \, .
\end{align}
\end{subequations}
The particle-number operator 
\begin{eqnarray}
A &\equiv& \sum_{p} c^{\dagger}_p c_p \, 
\end{eqnarray}
is thus obtained as a particular case with $o_{pq}\equiv \delta_{pq}$.

\section{Quasilinear form of BCCSD}
\label{quasilinearBCCSD}

\subsection{Definition of intermediates}

For both computational efficiency and simplicity of expression, it is useful to rewrite the nonlinear equations of BCCSD into quasilinear equations, in which each term consists of a single quasiparticle amplitude connected to an intermediate.  While these intermediates can be obtained from a diagrammatic procedure involving the similarity-transformed grand canonical potential (in connection to the coupled cluster effective-Hamiltonian diagrams derived in \textcite{shavitt09a}), a more straight-forward procedure will provide greater flexibility in the definition of the intermediates. As seen in the single-excitation amplitude equations of BCCSD, Eq.~\eqref{e:bccsd2qpex}, there are many terms which are nonlinear in $\mathcal{T}$, i.e. which contain more than one quasiparticle amplitude.  However, in each contribution consisting of multiple quasiparticle amplitudes, at least one of the quasiparticle amplitudes has $m$ external lines ($m \geq 1$).  There always exists a linear term containing the same quasiparticle amplitude with $m$ external lines, obtained at most through a renaming of indices.  For example, the sixth term of Eq. \eqref{e:bccsd2qpex} is the first nonlinear term, and one identifies immediately two quasiparticle amplitudes with external indices, $t_{\alpha k_1}$ and  $t_{k_2 \beta}$.  Both amplitudes are present as linear contributions, from the third term and second term, respectively (i.e., the terms involving $\Omega^{11}_{\beta k_1}$ and $\Omega^{11}_{\alpha k_1}$, respectively, where the second requires a renaming of index $k_1 \rightarrow k_2$). To connect the sixth diagram to the third diagram, one should ``integrate over'' the summation index $k_2$ to produce an intermediate $I$ with the desired indices $(\beta, k_1)$, i.e. one should rewrite the sixth term as an intermediate $I_{\beta k_1}$ connected to $t_{\alpha k_1}$.  Similarly, three later instances of $t_{\alpha k_1}$ can be found in Eq.~\eqref{e:bccsd2qpex} and can be manipulated in the same way to produce the full intermediate $\chi^{11}_{\beta k_1}$.  

\subsection{Amplitude equations with intermediates}

Implementing intermediates $\chi^{ij}$ and making use of permutation operators, the BCCSD amplitude equations from Eqs.~\eqref{e:bccsd2qpex} and~\eqref{e:bccsd4qpex} can be rewritten

\begin{widetext}
\begin{subequations}
\begin{align}
0 &= \Omega^{20}_{\alpha \beta} + P(\alpha/\beta) \displaystyle\sum_{k_1} \chi^{11}_{\beta k_1} t_{\alpha k_1}
 + \frac{1}{2}\sum_{k_1 k_2} \Big[ \Omega^{22}_{\alpha \beta k_1 k_2} t_{k_1 k_2} + \chi^{02}_{k_1 k_2} 
t_{\alpha \beta k_1 k_2} \Big] +\frac{1}{6} P(\alpha/\beta) \displaystyle\sum_{k_1 k_2 k_3} \Omega^{13}_{\alpha k_1 k_2 k_3}
 t_{k_1 k_2 k_3 \beta} \, , \\
0 &= \Omega^{40}_{\alpha \beta \gamma \delta} + P(\alpha \beta \gamma / \delta) \displaystyle\sum_{k_1} 
\Big[ \chi^{31}_{\alpha \beta \gamma k_1} t_{k_1 \delta} + 
\chi^{11\text{a}}_{\delta k_1} \; t_{\alpha \beta \gamma k_1} \Big] + \frac{1}{2} P(\alpha \beta / \gamma \delta) \displaystyle\sum_{k_1 k_2} \chi^{22}_{\alpha \beta k_1 k_2} 
t_{k_1 k_2 \gamma \delta}  \, ,
\end{align}
\end{subequations}
\end{widetext}
with the introduction of two separate intermediates $\chi^{11}$ and $\chi^{11\text{a}}$ due to the fact that the single-excitation amplitude equations and 
double-excitation amplitude equations have different factors from their respective number of identical $\mathcal{T}_m$ operators.  The intermediates are
defined as
\begin{widetext}
\begin{subequations}
\begin{align}
\chi^{02}_{k_1 k_2} &= \Omega^{02}_{k_1 k_2} + \frac{1}{2} \displaystyle\sum_{k_3 k_4} t_{k_3 k_4} \Omega^{04}_{k_3 k_4 k_1 k_2} \, , \\
\chi^{11}_{k_1 k_2} &= \Omega^{11}_{k_1 k_2} + \frac{1}{2} \displaystyle\sum_{k_3} t_{k_3 k_1} \Omega^{02}_{k_2 k_3} + \frac{1}{2}
 \displaystyle\sum_{k_3 k_4} t_{k_3 k_4} \Omega^{13}_{k_1 k_2 k_3 k_4} + \frac{1}{12} \displaystyle\sum_{k_3 k_4 k_5}  
\Omega^{04}_{k_2 k_3 k_4 k_5} \big( 3 t_{k_3 k_1} t_{k_4 k_5} + 2 t_{k_3 k_4 k_5 k_1} \big) \, , \\
\chi^{11\text{a}}_{k_1 k_2} &= \chi^{11}_{k_1 k_2} + \frac{1}{2} \displaystyle\sum_{k_3} t_{k_3 k_1} \Omega^{02}_{k_2 k_3} + 
\frac{1}{4}  \displaystyle\sum_{k_3 k_4 k_5} \Omega^{04}_{k_2 k_3 k_4 k_5} t_{k_3 k_1} t_{k_4 k_5} \, , \\
\chi^{22}_{k_1 k_2 k_3 k_4} &=  \Omega^{22}_{k_1 k_2 k_3 k_4} + \frac{1}{4} \displaystyle\sum_{k_5 k_6}  \Omega^{04}_{k_3 k_4 k_5 k_6} 
\big[ t_{k_5 k_6 k_1 k_2} - 4 t_{k_5 k_1} t_{k_6 k_2} \big] + 
P(k_1 / k_2) \displaystyle\sum_{k_5}  \Omega^{13}_{k_1 k_3 k_4 k_5} t_{k_5 k_2} \, , \\
\nonumber 
\chi^{31}_{k_1 k_2 k_3 k_4} &=  \Omega^{31}_{k_1 k_2 k_3 k_4} +  \frac{1}{2} P(k_1 / k_2 k_3) \displaystyle\sum_{k_5} 
\Omega^{22}_{k_2 k_3 k_5 k_4} t_{k_1 k_5} - \frac{1}{3} P(k_2 / k_1 k_3) \displaystyle\sum_{k_5 k_6}  \Omega^{13}_{k_2 k_5 k_6 k_4} 
t_{k_1 k_5} t_{k_6 k_3} \\ 
&+ \frac{1}{4} \displaystyle\sum_{k_5 k_6 k_7} \Omega^{04}_{k_5 k_6 k_7 k_4} 
t_{k_1 k_5} t_{k_2 k_6} t_{k_7 k_3}  \, .
\end{align}
\end{subequations}
\end{widetext}

\subsection{Diagrammatic method with intermediates}

\begin{figure}
\includegraphics[width=\columnwidth]{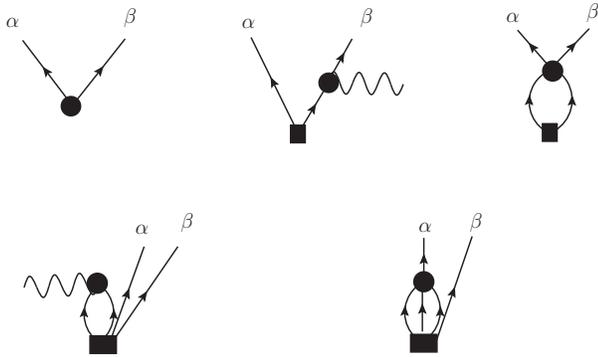}
\caption{Diagrammatic representation of the single-excitation amplitude equations in the quasilinear BCCSD approximation.}
\label{f:t1linear}
\end{figure}

\begin{figure}
\includegraphics[width=\columnwidth]{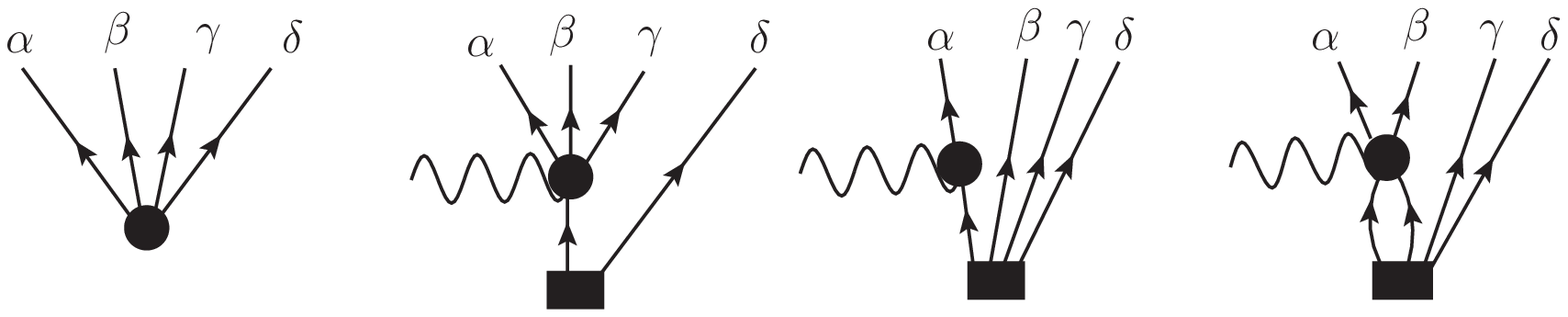}
\caption{Diagrammatic representation of the double-excitation amplitude equations in the quasilinear BCCSD approximation.}
\label{f:t2linear}
\end{figure}

\begin{figure*}
\includegraphics[width=\textwidth]{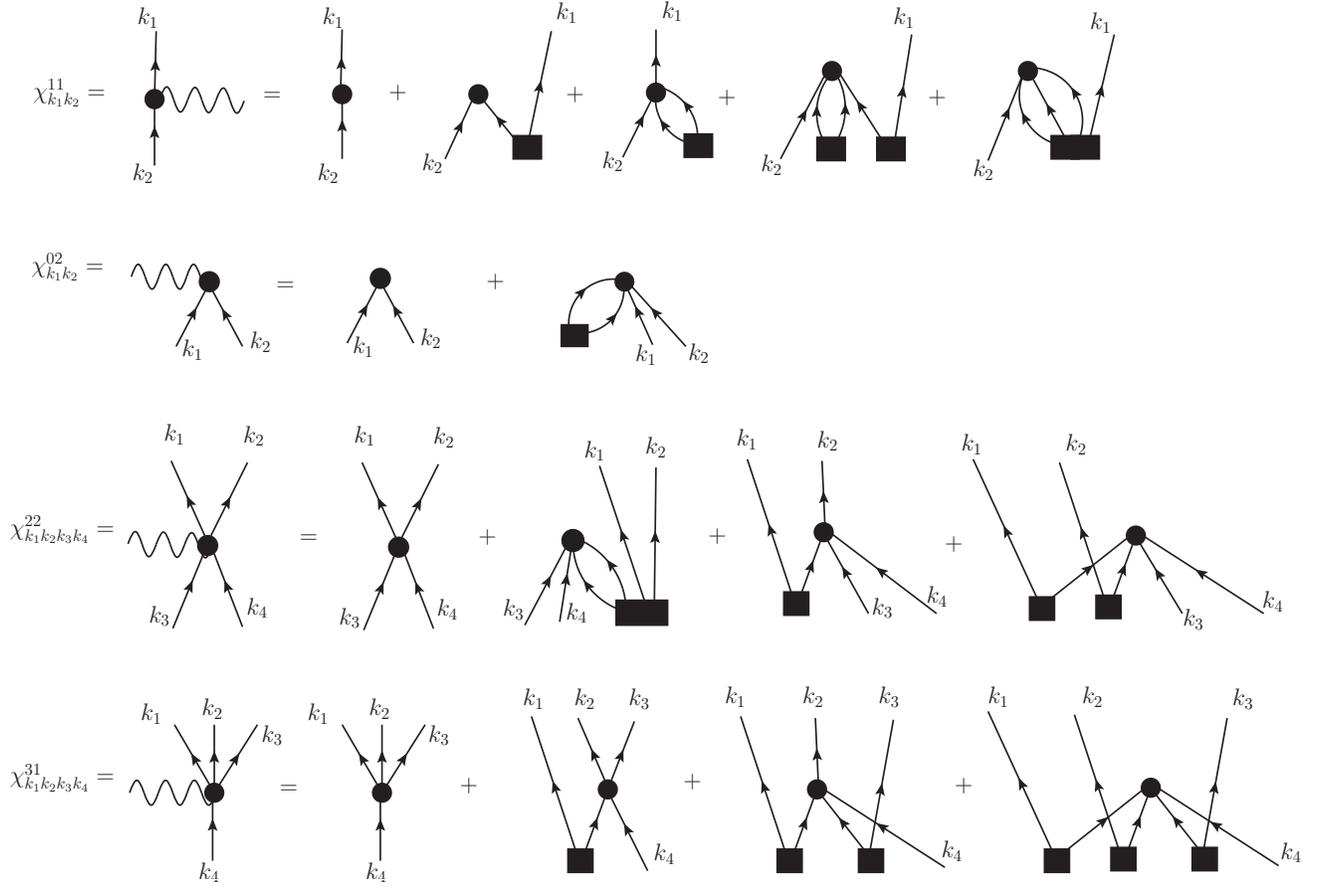}
\caption{Diagrammatic representation of the intermediates which enter into the amplitude equations of the quasilinear BCCSD approximation.  
Note that $\chi^{11}_{k_1 k_2}$ and $\chi^{11\text{a}}_{k_1 k_2}$ have an identical form diagrammatically, but their symmetry factors (i.e. the 
factor in front of the diagram) are different.  To differentiate the two, $\chi^{11}_{k_1 k_2}$ is drawn with a squiggly line to the right as shown here, 
while $\chi^{11\text{a}}_{k_1 k_2}$ has a squiggly line to the left as in Fig. \ref{f:t2linear}.}
\label{f:intdefs}
\end{figure*}

As before, the BCCSD equations, now in their quasilinear form, can be re-expressed in terms of diagrams to simplify and shorten their treatment.  It 
must be emphasized that the diagrammatic rules as developed in Section \ref{diagrammatic} do not apply directly to the intermediates and quasilinear 
form of BCCSD separately, as the symmetry factors can be affected when the full diagram is split.  Therefore, the diagrams with intermediates should 
truly be seen as a shorthand, while the original diagram should be employed to determine the corresponding algebraic expressions.  The 
single-excitation and double-excitation amplitude equations from Figs. \ref{f:bccsdt1} and \ref{f:bccsdt2} can be re-expressed in the simplified 
quasilinear form of Figs. \ref{f:t1linear} and \ref{f:t2linear}.  The definition of intermediates utilized in these figures is given in Fig. \ref{f:intdefs}.

\end{appendix}

\bibliography{bccsub}

\begin{thebibliography}{50}
\expandafter\ifx\csname natexlab\endcsname\relax\def\natexlab#1{#1}\fi
\expandafter\ifx\csname bibnamefont\endcsname\relax
  \def\bibnamefont#1{#1}\fi
\expandafter\ifx\csname bibfnamefont\endcsname\relax
  \def\bibfnamefont#1{#1}\fi
\expandafter\ifx\csname citenamefont\endcsname\relax
  \def\citenamefont#1{#1}\fi
\expandafter\ifx\csname url\endcsname\relax
  \def\url#1{\texttt{#1}}\fi
\expandafter\ifx\csname urlprefix\endcsname\relax\def\urlprefix{URL }\fi
\providecommand{\bibinfo}[2]{#2}
\providecommand{\eprint}[2][]{\url{#2}}

\bibitem[{\citenamefont{Dean and Hjorth-Jensen}(2004)}]{Dean:2003vc}
\bibinfo{author}{\bibfnamefont{D.}~\bibnamefont{Dean}} \bibnamefont{and}
  \bibinfo{author}{\bibfnamefont{M.}~\bibnamefont{Hjorth-Jensen}},
  \bibinfo{journal}{Phys. Rev.} \textbf{\bibinfo{volume}{C69}},
  \bibinfo{pages}{054320} (\bibinfo{year}{2004}).

\bibitem[{\citenamefont{Kowalski et~al.}(2004)\citenamefont{Kowalski, Dean,
  Hjorth-Jensen, Papenbrock, and Piecuch}}]{Kowalski:2003hp}
\bibinfo{author}{\bibfnamefont{K.}~\bibnamefont{Kowalski}},
  \bibinfo{author}{\bibfnamefont{D.}~\bibnamefont{Dean}},
  \bibinfo{author}{\bibfnamefont{M.}~\bibnamefont{Hjorth-Jensen}},
  \bibinfo{author}{\bibfnamefont{T.}~\bibnamefont{Papenbrock}},
  \bibnamefont{and} \bibinfo{author}{\bibfnamefont{P.}~\bibnamefont{Piecuch}},
  \bibinfo{journal}{Phys. Rev. Lett.} \textbf{\bibinfo{volume}{92}},
  \bibinfo{pages}{132501} (\bibinfo{year}{2004}).

\bibitem[{\citenamefont{Wloch et~al.}(2005{\natexlab{a}})\citenamefont{Wloch,
  Dean, Gour, Hjorth-Jensen, Kowalski et~al.}}]{Wloch:2005za}
\bibinfo{author}{\bibfnamefont{M.}~\bibnamefont{Wloch}},
  \bibinfo{author}{\bibfnamefont{D.}~\bibnamefont{Dean}},
  \bibinfo{author}{\bibfnamefont{J.}~\bibnamefont{Gour}},
  \bibinfo{author}{\bibfnamefont{M.}~\bibnamefont{Hjorth-Jensen}},
  \bibinfo{author}{\bibfnamefont{K.}~\bibnamefont{Kowalski}},
  \bibnamefont{et~al.}, \bibinfo{journal}{Phys. Rev. Lett.}
  \textbf{\bibinfo{volume}{94}}, \bibinfo{pages}{212501}
  (\bibinfo{year}{2005}{\natexlab{a}}).

\bibitem[{\citenamefont{Wloch et~al.}(2005{\natexlab{b}})\citenamefont{Wloch,
  Gour, Piecuch, Dean, Hjorth-Jensen et~al.}}]{Wloch:2005qq}
\bibinfo{author}{\bibfnamefont{M.}~\bibnamefont{Wloch}},
  \bibinfo{author}{\bibfnamefont{J.}~\bibnamefont{Gour}},
  \bibinfo{author}{\bibfnamefont{P.}~\bibnamefont{Piecuch}},
  \bibinfo{author}{\bibfnamefont{D.}~\bibnamefont{Dean}},
  \bibinfo{author}{\bibfnamefont{M.}~\bibnamefont{Hjorth-Jensen}},
  \bibnamefont{et~al.}, \bibinfo{journal}{J. Phys.}
  \textbf{\bibinfo{volume}{G31}}, \bibinfo{pages}{S1291}
  (\bibinfo{year}{2005}{\natexlab{b}}).

\bibitem[{\citenamefont{Gour et~al.}(2006)\citenamefont{Gour, Piecuch,
  Hjorth-Jensen, Wloch, and Dean}}]{Gour:2005dm}
\bibinfo{author}{\bibfnamefont{J.}~\bibnamefont{Gour}},
  \bibinfo{author}{\bibfnamefont{P.}~\bibnamefont{Piecuch}},
  \bibinfo{author}{\bibfnamefont{M.}~\bibnamefont{Hjorth-Jensen}},
  \bibinfo{author}{\bibfnamefont{M.}~\bibnamefont{Wloch}}, \bibnamefont{and}
  \bibinfo{author}{\bibfnamefont{D.}~\bibnamefont{Dean}},
  \bibinfo{journal}{Phys. Rev.} \textbf{\bibinfo{volume}{C74}},
  \bibinfo{pages}{024310} (\bibinfo{year}{2006}).

\bibitem[{\citenamefont{Hagen et~al.}(2007)\citenamefont{Hagen, Papenbrock,
  Dean, Schwenk, Nogga et~al.}}]{Hagen:2007ew}
\bibinfo{author}{\bibfnamefont{G.}~\bibnamefont{Hagen}},
  \bibinfo{author}{\bibfnamefont{T.}~\bibnamefont{Papenbrock}},
  \bibinfo{author}{\bibfnamefont{D.}~\bibnamefont{Dean}},
  \bibinfo{author}{\bibfnamefont{A.}~\bibnamefont{Schwenk}},
  \bibinfo{author}{\bibfnamefont{A.}~\bibnamefont{Nogga}},
  \bibnamefont{et~al.}, \bibinfo{journal}{Phys. Rev.}
  \textbf{\bibinfo{volume}{C76}}, \bibinfo{pages}{034302}
  (\bibinfo{year}{2007}).

\bibitem[{\citenamefont{Hagen et~al.}(2010)\citenamefont{Hagen, Papenbrock,
  Dean, and Hjorth-Jensen}}]{Hagen:2010gd}
\bibinfo{author}{\bibfnamefont{G.}~\bibnamefont{Hagen}},
  \bibinfo{author}{\bibfnamefont{T.}~\bibnamefont{Papenbrock}},
  \bibinfo{author}{\bibfnamefont{D.~J.} \bibnamefont{Dean}}, \bibnamefont{and}
  \bibinfo{author}{\bibfnamefont{M.}~\bibnamefont{Hjorth-Jensen}},
  \bibinfo{journal}{Phys. Rev.} \textbf{\bibinfo{volume}{C82}},
  \bibinfo{pages}{034330} (\bibinfo{year}{2010}).

\bibitem[{\citenamefont{Jansen et~al.}(2011)\citenamefont{Jansen,
  Hjorth-Jensen, Hagen, and Papenbrock}}]{Jansen:2011gb}
\bibinfo{author}{\bibfnamefont{G.}~\bibnamefont{Jansen}},
  \bibinfo{author}{\bibfnamefont{M.}~\bibnamefont{Hjorth-Jensen}},
  \bibinfo{author}{\bibfnamefont{G.}~\bibnamefont{Hagen}}, \bibnamefont{and}
  \bibinfo{author}{\bibfnamefont{T.}~\bibnamefont{Papenbrock}},
  \bibinfo{journal}{Phys. Rev.} \textbf{\bibinfo{volume}{C83}},
  \bibinfo{pages}{054306} (\bibinfo{year}{2011}).

\bibitem[{\citenamefont{Binder et~al.}(2013{\natexlab{a}})\citenamefont{Binder,
  Langhammer, Calci, Navratil, and Roth}}]{Binder:2012mk}
\bibinfo{author}{\bibfnamefont{S.}~\bibnamefont{Binder}},
  \bibinfo{author}{\bibfnamefont{J.}~\bibnamefont{Langhammer}},
  \bibinfo{author}{\bibfnamefont{A.}~\bibnamefont{Calci}},
  \bibinfo{author}{\bibfnamefont{P.}~\bibnamefont{Navratil}}, \bibnamefont{and}
  \bibinfo{author}{\bibfnamefont{R.}~\bibnamefont{Roth}},
  \bibinfo{journal}{Phys. Rev.} \textbf{\bibinfo{volume}{C87}},
  \bibinfo{pages}{021303} (\bibinfo{year}{2013}{\natexlab{a}}).

\bibitem[{\citenamefont{Binder et~al.}(2013{\natexlab{b}})\citenamefont{Binder,
  Piecuch, Calci, Langhammer, Navrátil et~al.}}]{Binder:2013oea}
\bibinfo{author}{\bibfnamefont{S.}~\bibnamefont{Binder}},
  \bibinfo{author}{\bibfnamefont{P.}~\bibnamefont{Piecuch}},
  \bibinfo{author}{\bibfnamefont{A.}~\bibnamefont{Calci}},
  \bibinfo{author}{\bibfnamefont{J.}~\bibnamefont{Langhammer}},
  \bibinfo{author}{\bibfnamefont{P.}~\bibnamefont{Navrátil}},
  \bibnamefont{et~al.}, \bibinfo{journal}{Phys.Rev.}
  \textbf{\bibinfo{volume}{C88}}, \bibinfo{pages}{054319}
  (\bibinfo{year}{2013}{\natexlab{b}}).

\bibitem[{\citenamefont{Barbieri and Dickhoff}(2001)}]{Barbieri:2000pg}
\bibinfo{author}{\bibfnamefont{C.}~\bibnamefont{Barbieri}} \bibnamefont{and}
  \bibinfo{author}{\bibfnamefont{W.}~\bibnamefont{Dickhoff}},
  \bibinfo{journal}{Phys. Rev.} \textbf{\bibinfo{volume}{C63}},
  \bibinfo{pages}{034313} (\bibinfo{year}{2001}).

\bibitem[{\citenamefont{Barbieri and Dickhoff}(2002)}]{Barbieri:2001gt}
\bibinfo{author}{\bibfnamefont{C.}~\bibnamefont{Barbieri}} \bibnamefont{and}
  \bibinfo{author}{\bibfnamefont{W.}~\bibnamefont{Dickhoff}},
  \bibinfo{journal}{Phys. Rev.} \textbf{\bibinfo{volume}{C65}},
  \bibinfo{pages}{064313} (\bibinfo{year}{2002}).

\bibitem[{\citenamefont{Dickhoff and Barbieri}(2004)}]{Dickhoff:2004xx}
\bibinfo{author}{\bibfnamefont{W.~H.} \bibnamefont{Dickhoff}} \bibnamefont{and}
  \bibinfo{author}{\bibfnamefont{C.}~\bibnamefont{Barbieri}},
  \bibinfo{journal}{Prog. Part. Nucl. Phys.} \textbf{\bibinfo{volume}{52}},
  \bibinfo{pages}{377} (\bibinfo{year}{2004}).

\bibitem[{\citenamefont{Waldecker et~al.}(2011)\citenamefont{Waldecker,
  Barbieri, and Dickhoff}}]{Waldecker:2011by}
\bibinfo{author}{\bibfnamefont{S.}~\bibnamefont{Waldecker}},
  \bibinfo{author}{\bibfnamefont{C.}~\bibnamefont{Barbieri}}, \bibnamefont{and}
  \bibinfo{author}{\bibfnamefont{W.}~\bibnamefont{Dickhoff}},
  \bibinfo{journal}{Phys. Rev.} \textbf{\bibinfo{volume}{C84}},
  \bibinfo{pages}{034616} (\bibinfo{year}{2011}).

\bibitem[{\citenamefont{Cipollone et~al.}(2013)\citenamefont{Cipollone,
  Barbieri, and Navrátil}}]{Cipollone:2013zma}
\bibinfo{author}{\bibfnamefont{A.}~\bibnamefont{Cipollone}},
  \bibinfo{author}{\bibfnamefont{C.}~\bibnamefont{Barbieri}}, \bibnamefont{and}
  \bibinfo{author}{\bibfnamefont{P.}~\bibnamefont{Navrátil}},
  \bibinfo{journal}{Phys. Rev. Lett.} \textbf{\bibinfo{volume}{11}},
  \bibinfo{pages}{062501} (\bibinfo{year}{2013}).

\bibitem[{\citenamefont{Tsukiyama et~al.}(2011)\citenamefont{Tsukiyama, Bogner,
  and Schwenk}}]{Tsukiyama:2010rj}
\bibinfo{author}{\bibfnamefont{K.}~\bibnamefont{Tsukiyama}},
  \bibinfo{author}{\bibfnamefont{S.}~\bibnamefont{Bogner}}, \bibnamefont{and}
  \bibinfo{author}{\bibfnamefont{A.}~\bibnamefont{Schwenk}},
  \bibinfo{journal}{Phys. Rev. Lett.} \textbf{\bibinfo{volume}{106}},
  \bibinfo{pages}{222502} (\bibinfo{year}{2011}).

\bibitem[{\citenamefont{Hergert
  et~al.}(2013{\natexlab{a}})\citenamefont{Hergert, Bogner, Binder, Calci,
  Langhammer, Roth, and Schwenk}}]{Hergert:2012nb}
\bibinfo{author}{\bibfnamefont{H.}~\bibnamefont{Hergert}},
  \bibinfo{author}{\bibfnamefont{S.}~\bibnamefont{Bogner}},
  \bibinfo{author}{\bibfnamefont{S.}~\bibnamefont{Binder}},
  \bibinfo{author}{\bibfnamefont{A.}~\bibnamefont{Calci}},
  \bibinfo{author}{\bibfnamefont{J.}~\bibnamefont{Langhammer}},
  \bibinfo{author}{\bibfnamefont{R.}~\bibnamefont{Roth}}, \bibnamefont{and}
  \bibinfo{author}{\bibfnamefont{A.}~\bibnamefont{Schwenk}},
  \bibinfo{journal}{Phys. Rev.} \textbf{\bibinfo{volume}{C87}},
  \bibinfo{pages}{034307} (\bibinfo{year}{2013}{\natexlab{a}}).

\bibitem[{\citenamefont{Binder et~al.}(2014)\citenamefont{Binder, Langhammer,
  Calci, and Roth}}]{Binder:2013xaa}
\bibinfo{author}{\bibfnamefont{S.}~\bibnamefont{Binder}},
  \bibinfo{author}{\bibfnamefont{J.}~\bibnamefont{Langhammer}},
  \bibinfo{author}{\bibfnamefont{A.}~\bibnamefont{Calci}}, \bibnamefont{and}
  \bibinfo{author}{\bibfnamefont{R.}~\bibnamefont{Roth}},
  \bibinfo{journal}{Phys.Lett.} \textbf{\bibinfo{volume}{B736}},
  \bibinfo{pages}{119} (\bibinfo{year}{2014}).

\bibitem[{\citenamefont{Hergert
  et~al.}(2013{\natexlab{b}})\citenamefont{Hergert, Binder, Calci, Langhammer,
  and Roth}}]{Hergert:2013uja}
\bibinfo{author}{\bibfnamefont{H.}~\bibnamefont{Hergert}},
  \bibinfo{author}{\bibfnamefont{S.}~\bibnamefont{Binder}},
  \bibinfo{author}{\bibfnamefont{A.}~\bibnamefont{Calci}},
  \bibinfo{author}{\bibfnamefont{J.}~\bibnamefont{Langhammer}},
  \bibnamefont{and} \bibinfo{author}{\bibfnamefont{R.}~\bibnamefont{Roth}},
  \bibinfo{journal}{Phys. Rev. Lett.} \textbf{\bibinfo{volume}{110}},
  \bibinfo{pages}{242501} (\bibinfo{year}{2013}{\natexlab{b}}).

\bibitem[{\citenamefont{Jansen et~al.}(2014)\citenamefont{Jansen, Engel, Hagen,
  Navratil, and Signoracci}}]{Jansen:2014qxa}
\bibinfo{author}{\bibfnamefont{G.}~\bibnamefont{Jansen}},
  \bibinfo{author}{\bibfnamefont{J.}~\bibnamefont{Engel}},
  \bibinfo{author}{\bibfnamefont{G.}~\bibnamefont{Hagen}},
  \bibinfo{author}{\bibfnamefont{P.}~\bibnamefont{Navratil}}, \bibnamefont{and}
  \bibinfo{author}{\bibfnamefont{A.}~\bibnamefont{Signoracci}}
  (\bibinfo{year}{2014}), \bibinfo{note}{arXiv:1402.2563}.

\bibitem[{\citenamefont{Bogner et~al.}(2014)\citenamefont{Bogner, Hergert,
  Holt, Schwenk, Binder et~al.}}]{Bogner:2014baa}
\bibinfo{author}{\bibfnamefont{S.}~\bibnamefont{Bogner}},
  \bibinfo{author}{\bibfnamefont{H.}~\bibnamefont{Hergert}},
  \bibinfo{author}{\bibfnamefont{J.}~\bibnamefont{Holt}},
  \bibinfo{author}{\bibfnamefont{A.}~\bibnamefont{Schwenk}},
  \bibinfo{author}{\bibfnamefont{S.}~\bibnamefont{Binder}},
  \bibnamefont{et~al.} (\bibinfo{year}{2014}), \bibinfo{note}{arXiv:1402.1407}.

\bibitem[{\citenamefont{Som\`a et~al.}(2011)\citenamefont{Som\`a, Duguet, and
  Barbieri}}]{soma11a}
\bibinfo{author}{\bibfnamefont{V.}~\bibnamefont{Som\`a}},
  \bibinfo{author}{\bibfnamefont{T.}~\bibnamefont{Duguet}}, \bibnamefont{and}
  \bibinfo{author}{\bibfnamefont{C.}~\bibnamefont{Barbieri}},
  \bibinfo{journal}{Phys. Rev.} \textbf{\bibinfo{volume}{C84}},
  \bibinfo{pages}{064317} (\bibinfo{year}{2011}).

\bibitem[{\citenamefont{Som\`a et~al.}(2013)\citenamefont{Som\`a, Barbieri, and
  Duguet}}]{Soma:2012zd}
\bibinfo{author}{\bibfnamefont{V.}~\bibnamefont{Som\`a}},
  \bibinfo{author}{\bibfnamefont{C.}~\bibnamefont{Barbieri}}, \bibnamefont{and}
  \bibinfo{author}{\bibfnamefont{T.}~\bibnamefont{Duguet}},
  \bibinfo{journal}{Phys. Rev.} \textbf{\bibinfo{volume}{C87}},
  \bibinfo{pages}{011303} (\bibinfo{year}{2013}).

\bibitem[{\citenamefont{Barbieri et~al.}(2012)\citenamefont{Barbieri,
  Cipollone, Som\`a, Duguet, and Navratil}}]{Barbieri:2012rd}
\bibinfo{author}{\bibfnamefont{C.}~\bibnamefont{Barbieri}},
  \bibinfo{author}{\bibfnamefont{A.}~\bibnamefont{Cipollone}},
  \bibinfo{author}{\bibfnamefont{V.}~\bibnamefont{Som\`a}},
  \bibinfo{author}{\bibfnamefont{T.}~\bibnamefont{Duguet}}, \bibnamefont{and}
  \bibinfo{author}{\bibfnamefont{P.}~\bibnamefont{Navratil}}
  (\bibinfo{year}{2012}), \bibinfo{note}{arXiv:1211.3315}.

\bibitem[{\citenamefont{Soma et~al.}(2014)\citenamefont{Soma, Barbieri,
  Cipollone, Duguet, and Navratil}}]{Soma:2013vca}
\bibinfo{author}{\bibfnamefont{V.}~\bibnamefont{Soma}},
  \bibinfo{author}{\bibfnamefont{C.}~\bibnamefont{Barbieri}},
  \bibinfo{author}{\bibfnamefont{A.}~\bibnamefont{Cipollone}},
  \bibinfo{author}{\bibfnamefont{T.}~\bibnamefont{Duguet}}, \bibnamefont{and}
  \bibinfo{author}{\bibfnamefont{P.}~\bibnamefont{Navratil}},
  \bibinfo{journal}{EPJ Web Conf.} \textbf{\bibinfo{volume}{66}},
  \bibinfo{pages}{02005} (\bibinfo{year}{2014}).

\bibitem[{\citenamefont{Somà et~al.}(2014)\citenamefont{Somà, Cipollone,
  Barbieri, Navrátil, and Duguet}}]{Soma:2013xha}
\bibinfo{author}{\bibfnamefont{V.}~\bibnamefont{Somà}},
  \bibinfo{author}{\bibfnamefont{A.}~\bibnamefont{Cipollone}},
  \bibinfo{author}{\bibfnamefont{C.}~\bibnamefont{Barbieri}},
  \bibinfo{author}{\bibfnamefont{P.}~\bibnamefont{Navrátil}},
  \bibnamefont{and} \bibinfo{author}{\bibfnamefont{T.}~\bibnamefont{Duguet}},
  \bibinfo{journal}{Phys.Rev.} \textbf{\bibinfo{volume}{C89}},
  \bibinfo{pages}{061301} (\bibinfo{year}{2014}).

\bibitem[{\citenamefont{Stolarczyk and Monkhorst}(2010)}]{stolarczyk2010}
\bibinfo{author}{\bibfnamefont{L.}~\bibnamefont{Stolarczyk}} \bibnamefont{and}
  \bibinfo{author}{\bibfnamefont{H.}~\bibnamefont{Monkhorst}},
  \bibinfo{journal}{Mol. Phys.} \textbf{\bibinfo{volume}{108}},
  \bibinfo{pages}{3067} (\bibinfo{year}{2010}).

\bibitem[{\citenamefont{Emrich and Zabolitzky}(1984)}]{emrich84a}
\bibinfo{author}{\bibfnamefont{K.}~\bibnamefont{Emrich}} \bibnamefont{and}
  \bibinfo{author}{\bibfnamefont{J.~G.} \bibnamefont{Zabolitzky}},
  \bibinfo{journal}{Phys. Rev.} \textbf{\bibinfo{volume}{B30}},
  \bibinfo{pages}{2049} (\bibinfo{year}{1984}).

\bibitem[{\citenamefont{Lahoz and Bishop}(1988)}]{lahoz88a}
\bibinfo{author}{\bibfnamefont{W.~A.} \bibnamefont{Lahoz}} \bibnamefont{and}
  \bibinfo{author}{\bibfnamefont{R.~F.} \bibnamefont{Bishop}},
  \bibinfo{journal}{Z. Phys} \textbf{\bibinfo{volume}{B73}},
  \bibinfo{pages}{363} (\bibinfo{year}{1988}).

\bibitem[{\citenamefont{Henderson et~al.}(2014)\citenamefont{Henderson,
  Scuseria, Dukelsky, Signoracci, and Duguet}}]{Henderson:2014vka}
\bibinfo{author}{\bibfnamefont{T.~M.} \bibnamefont{Henderson}},
  \bibinfo{author}{\bibfnamefont{G.~E.} \bibnamefont{Scuseria}},
  \bibinfo{author}{\bibfnamefont{J.}~\bibnamefont{Dukelsky}},
  \bibinfo{author}{\bibfnamefont{A.}~\bibnamefont{Signoracci}},
  \bibnamefont{and} \bibinfo{author}{\bibfnamefont{T.}~\bibnamefont{Duguet}},
  \bibinfo{journal}{Phys. Rev.} \textbf{\bibinfo{volume}{C89}},
  \bibinfo{pages}{054305} (\bibinfo{year}{2014}).

\bibitem[{\citenamefont{Richardson}(1963)}]{richardson63a}
\bibinfo{author}{\bibfnamefont{R.~W.} \bibnamefont{Richardson}},
  \bibinfo{journal}{Phys. Lett.} \textbf{\bibinfo{volume}{3}},
  \bibinfo{pages}{277} (\bibinfo{year}{1963}).

\bibitem[{\citenamefont{Richardson and Sherman}(1964)}]{richardson64a}
\bibinfo{author}{\bibfnamefont{R.~W.} \bibnamefont{Richardson}}
  \bibnamefont{and} \bibinfo{author}{\bibfnamefont{N.}~\bibnamefont{Sherman}},
  \bibinfo{journal}{Nucl. Phys.} \textbf{\bibinfo{volume}{52}},
  \bibinfo{pages}{221} (\bibinfo{year}{1964}).

\bibitem[{\citenamefont{Jansen}(2013)}]{Jansen:2013gb}
\bibinfo{author}{\bibfnamefont{G.}~\bibnamefont{Jansen}},
  \bibinfo{journal}{Phys. Rev.} \textbf{\bibinfo{volume}{C88}},
  \bibinfo{pages}{024305} (\bibinfo{year}{2013}).

\bibitem[{\citenamefont{Hergert and Roth}(2009)}]{Hergert:2009na}
\bibinfo{author}{\bibfnamefont{H.}~\bibnamefont{Hergert}} \bibnamefont{and}
  \bibinfo{author}{\bibfnamefont{R.}~\bibnamefont{Roth}},
  \bibinfo{journal}{Phys. Lett.} \textbf{\bibinfo{volume}{B682}},
  \bibinfo{pages}{27} (\bibinfo{year}{2009}).

\bibitem[{\citenamefont{Ring and Schuck}(1980)}]{ring80a}
\bibinfo{author}{\bibfnamefont{P.}~\bibnamefont{Ring}} \bibnamefont{and}
  \bibinfo{author}{\bibfnamefont{P.}~\bibnamefont{Schuck}},
  \emph{\bibinfo{title}{The Nuclear Many-Body Problem}}
  (\bibinfo{publisher}{Springer-Verlag}, \bibinfo{address}{New-York},
  \bibinfo{year}{1980}).

\bibitem[{\citenamefont{Carbone et~al.}(2013)\citenamefont{Carbone, Cipollone,
  Barbieri, Rios, and Polls}}]{Carbone:2013eqa}
\bibinfo{author}{\bibfnamefont{A.}~\bibnamefont{Carbone}},
  \bibinfo{author}{\bibfnamefont{A.}~\bibnamefont{Cipollone}},
  \bibinfo{author}{\bibfnamefont{C.}~\bibnamefont{Barbieri}},
  \bibinfo{author}{\bibfnamefont{A.}~\bibnamefont{Rios}}, \bibnamefont{and}
  \bibinfo{author}{\bibfnamefont{A.}~\bibnamefont{Polls}},
  \bibinfo{journal}{Phys. Rev.} \textbf{\bibinfo{volume}{C88}},
  \bibinfo{pages}{054326} (\bibinfo{year}{2013}).

\bibitem[{\citenamefont{Handy et~al.}(1989)\citenamefont{Handy, Pople,
  Head-Gordon, Raghavachari, and Trucks}}]{handy89a}
\bibinfo{author}{\bibfnamefont{N.~C.} \bibnamefont{Handy}},
  \bibinfo{author}{\bibfnamefont{J.}~\bibnamefont{Pople}},
  \bibinfo{author}{\bibfnamefont{M.}~\bibnamefont{Head-Gordon}},
  \bibinfo{author}{\bibfnamefont{K.}~\bibnamefont{Raghavachari}},
  \bibnamefont{and} \bibinfo{author}{\bibfnamefont{G.~W.}
  \bibnamefont{Trucks}}, \bibinfo{journal}{Chem. Phys. Lett.}
  \textbf{\bibinfo{volume}{164}}, \bibinfo{pages}{185} (\bibinfo{year}{1989}).

\bibitem[{\citenamefont{Rotival and Duguet}(2009)}]{Rotival:2007hp}
\bibinfo{author}{\bibfnamefont{V.}~\bibnamefont{Rotival}} \bibnamefont{and}
  \bibinfo{author}{\bibfnamefont{T.}~\bibnamefont{Duguet}},
  \bibinfo{journal}{Phys. Rev.} \textbf{\bibinfo{volume}{C79}},
  \bibinfo{pages}{054308} (\bibinfo{year}{2009}).

\bibitem[{\citenamefont{Shavitt and Bartlett}(2009)}]{shavitt09a}
\bibinfo{author}{\bibfnamefont{I.}~\bibnamefont{Shavitt}} \bibnamefont{and}
  \bibinfo{author}{\bibfnamefont{R.~J.} \bibnamefont{Bartlett}},
  \emph{\bibinfo{title}{Many-Body Methods in Chemistry and Physics}}
  (\bibinfo{publisher}{Cambridge University Press}, \bibinfo{year}{2009}).

\bibitem[{\citenamefont{Taube and Bartlett}(2008)}]{taube08a}
\bibinfo{author}{\bibfnamefont{A.~G.} \bibnamefont{Taube}} \bibnamefont{and}
  \bibinfo{author}{\bibfnamefont{R.~J.} \bibnamefont{Bartlett}},
  \bibinfo{journal}{J. Chem. Phys.} \textbf{\bibinfo{volume}{128}},
  \bibinfo{pages}{044110} (\bibinfo{year}{2008}).

\bibitem[{\citenamefont{Piecuch and Wloch}(2005)}]{piecuch05a}
\bibinfo{author}{\bibfnamefont{P.}~\bibnamefont{Piecuch}} \bibnamefont{and}
  \bibinfo{author}{\bibfnamefont{M.}~\bibnamefont{Wloch}}, \bibinfo{journal}{J.
  Chem. Phys.} \textbf{\bibinfo{volume}{123}}, \bibinfo{pages}{224105}
  (\bibinfo{year}{2005}).

\bibitem[{\citenamefont{Duguet}(2014{\natexlab{a}})}]{duguet14a}
\bibinfo{author}{\bibfnamefont{T.}~\bibnamefont{Duguet}}
  (\bibinfo{year}{2014}{\natexlab{a}}), \bibinfo{note}{in preparation}.

\bibitem[{\citenamefont{Duguet}(2014{\natexlab{b}})}]{Duguet:2014jja}
\bibinfo{author}{\bibfnamefont{T.}~\bibnamefont{Duguet}}
  (\bibinfo{year}{2014}{\natexlab{b}}), \bibinfo{note}{arXiv:1406.7183}.

\bibitem[{\citenamefont{Ekstrom et~al.}(2013)\citenamefont{Ekstrom, Baardsen,
  Forssen, Hagen, Hjorth-Jensen, Jansen, Machleidt, Nazarewicz, Papenbrock,
  Sarich et~al.}}]{Ekstrom:2013kea}
\bibinfo{author}{\bibfnamefont{A.}~\bibnamefont{Ekstrom}},
  \bibinfo{author}{\bibfnamefont{G.}~\bibnamefont{Baardsen}},
  \bibinfo{author}{\bibfnamefont{C.}~\bibnamefont{Forssen}},
  \bibinfo{author}{\bibfnamefont{G.}~\bibnamefont{Hagen}},
  \bibinfo{author}{\bibfnamefont{M.}~\bibnamefont{Hjorth-Jensen}},
  \bibinfo{author}{\bibfnamefont{G.~R.} \bibnamefont{Jansen}},
  \bibinfo{author}{\bibfnamefont{R.}~\bibnamefont{Machleidt}},
  \bibinfo{author}{\bibfnamefont{W.}~\bibnamefont{Nazarewicz}},
  \bibinfo{author}{\bibfnamefont{T.}~\bibnamefont{Papenbrock}},
  \bibinfo{author}{\bibfnamefont{J.}~\bibnamefont{Sarich}},
  \bibnamefont{et~al.}, \bibinfo{journal}{Phys. Rev. Lett.}
  \textbf{\bibinfo{volume}{110}}, \bibinfo{pages}{192502}
  (\bibinfo{year}{2013}).

\bibitem[{\citenamefont{Shen and Piecuch}(2013)}]{piecuch2013}
\bibinfo{author}{\bibfnamefont{J.}~\bibnamefont{Shen}} \bibnamefont{and}
  \bibinfo{author}{\bibfnamefont{P.}~\bibnamefont{Piecuch}},
  \bibinfo{journal}{J. Chem. Phys.} \textbf{\bibinfo{volume}{138}},
  \bibinfo{pages}{194102} (\bibinfo{year}{2013}).

\bibitem[{\citenamefont{Thouless}(1961)}]{thouless61a}
\bibinfo{author}{\bibfnamefont{D.~J.} \bibnamefont{Thouless}},
  \bibinfo{journal}{Nucl. Phys.} \textbf{\bibinfo{volume}{21}},
  \bibinfo{pages}{225} (\bibinfo{year}{1961}).

\bibitem[{\citenamefont{Furnstahl
  et~al.}(2014{\natexlab{a}})\citenamefont{Furnstahl, More, and
  Papenbrock}}]{furnstahl14}
\bibinfo{author}{\bibfnamefont{R.}~\bibnamefont{Furnstahl}},
  \bibinfo{author}{\bibfnamefont{S.}~\bibnamefont{More}}, \bibnamefont{and}
  \bibinfo{author}{\bibfnamefont{T.}~\bibnamefont{Papenbrock}},
  \bibinfo{journal}{Phys. Rev. C} \textbf{\bibinfo{volume}{89}},
  \bibinfo{pages}{044301} (\bibinfo{year}{2014}{\natexlab{a}}).

\bibitem[{\citenamefont{Furnstahl
  et~al.}(2014{\natexlab{b}})\citenamefont{Furnstahl, Hagen, Papenbrock, and
  Wendt}}]{furnstahl14a}
\bibinfo{author}{\bibfnamefont{R.}~\bibnamefont{Furnstahl}},
  \bibinfo{author}{\bibfnamefont{G.}~\bibnamefont{Hagen}},
  \bibinfo{author}{\bibfnamefont{T.}~\bibnamefont{Papenbrock}},
  \bibnamefont{and} \bibinfo{author}{\bibfnamefont{K.}~\bibnamefont{Wendt}}
  (\bibinfo{year}{2014}{\natexlab{b}}), \eprint{arXiv:1408.0252}.

\bibitem[{\citenamefont{Wendt}(2014)}]{kylepc}
\bibinfo{author}{\bibfnamefont{K.}~\bibnamefont{Wendt}} (\bibinfo{year}{2014}),
  \bibinfo{note}{private communication}.

\bibitem[{\citenamefont{Wang et~al.}(2012)\citenamefont{Wang, Audi, Wapstra,
  Kondev, MacCormick, Xu, and Pfeiffer}}]{ame2012}
\bibinfo{author}{\bibfnamefont{M.}~\bibnamefont{Wang}},
  \bibinfo{author}{\bibfnamefont{G.}~\bibnamefont{Audi}},
  \bibinfo{author}{\bibfnamefont{A.~H.} \bibnamefont{Wapstra}},
  \bibinfo{author}{\bibfnamefont{F.}~\bibnamefont{Kondev}},
  \bibinfo{author}{\bibfnamefont{M.}~\bibnamefont{MacCormick}},
  \bibinfo{author}{\bibfnamefont{X.}~\bibnamefont{Xu}}, \bibnamefont{and}
  \bibinfo{author}{\bibfnamefont{B.}~\bibnamefont{Pfeiffer}},
  \bibinfo{journal}{Chin. Phys.} \textbf{\bibinfo{volume}{C36}},
  \bibinfo{pages}{1603} (\bibinfo{year}{2012}).

\end{thebibliography}

\end{document}